\documentclass{article}
\usepackage[margin=2cm]{geometry}
\usepackage{lscape}

\usepackage[utf8]{inputenc} 
\usepackage[T1]{fontenc}    
\usepackage{hyperref}       
\usepackage{url}            
\usepackage{booktabs}       
\usepackage{amsfonts}       
\usepackage{nicefrac}       
\usepackage{microtype}      
\usepackage{lipsum}		
\usepackage{graphicx}
\usepackage{cite}

\usepackage{array}
\usepackage{amssymb, amsmath, amsthm}
\usepackage{graphicx}
\usepackage{lmodern,url}
\usepackage{makecell} 
\usepackage{cancel}
\usepackage{multirow}
\usepackage{microtype}
\usepackage{lineno}
\usepackage{lscape}
\usepackage{xspace}
\usepackage{xcolor}
\usepackage{siunitx}
\usepackage{todonotes}
\usepackage{booktabs}
\usepackage[ruled,vlined]{algorithm2e}
\usepackage{optidef}
\usepackage{mathdots}
\usepackage{subcaption}
\usepackage{csquotes}
\usepackage{caption}
\usepackage{verbatim}
\usepackage{ulem}
\usepackage{amsmath}
\usepackage{tikz}
\usetikzlibrary{fit}
\usetikzlibrary{positioning}
\tikzset{%
  highlight/.style={rectangle,rounded corners,fill=red!15,draw,fill opacity=0.25,thick,inner sep=0pt}
}

\definecolor{mygreen}{RGB}{160, 242,182}

\definecolor{mypurple}{RGB}{236, 223, 234}

\definecolor{red}{rgb}{1,0,0}
\definecolor{green}{rgb}{0,1,0}
\definecolor{blue}{rgb}{0,0,1}

\usepackage{float}
\usepackage{siunitx}
\usepackage{subcaption}
\usepackage{tikz} 
\usetikzlibrary{positioning} 
\usetikzlibrary{calc} 

\usepackage{ mathrsfs }

\allowdisplaybreaks

\title{Synergistic Effects of Behavioral Feedback and Seasonality Generate Chaos in Cooperative Multi-Pathogen Systems}
\usepackage{authblk}

\author[1]{Rodrigo Amaral Lind}
\author[2]{Fakhteh Ghanbarnejad}
\author[3,4*]{Seba Contreras}

\affil[1]{Institute for the Dynamics of Complex Systems, University of G\"ottingen, G\"ottingen, Germany.}
\affil[2]{School of Technology and Architecture, SRH University of Applied Sciences Heidelberg, Leipzig Campus, Germany}
\affil[3]{Hangzhou International Innovation Institute, Beihang University, Hangzhou, China.}
\affil[4]{Interdisciplinary Center for the Mathematical Modeling of Infectious Disease Dynamics (IMMIDD), University of Münster, Münster, Germany.}

\affil[ ]{* Corresponding Author: Seba Contreras (contreras@buaa.edu.cn)}

\date{}

\newcommand{\A}{X_{IS}}
\newcommand{\B}{X_{SI}}
\newcommand{\aB}{X_{RI}}
\newcommand{\Ab}{X_{IR}}
\newcommand{\AB}{X_{II}}
\newcommand{\aVar}{X_{RS}}
\newcommand{\bb}{X_{SR}}
\newcommand{\ab}{X_{RR}}
\newcommand{\X}{I}
\newcommand{\XA}{I_A}
\newcommand{\XB}{I_B}

\begin{document}

\maketitle

\begin{abstract}
Infectious diseases may interact by competing for the same hosts or by facilitating subsequent infections. Understanding the dynamics of such multi-pathogen systems, particularly those subject to endemic seasonality and mitigation, is essential for designing robust public health interventions. We propose a three-stage modeling framework to disentangle the interplay between seasonality and behavioral feedback as a way of mitigation. First, we analyze a coupled susceptible-infectious-recovered-susceptible (SIRS) system without external forcing and show that the abrupt transition between the disease-free and endemic equilibria arises from a backward bifurcation-induced first-order phase transition. Second, we independently examine seasonality and behavioral feedback, characterizing where and when oscillatory behavior is induced near critical tipping points. Third, we demonstrate that their combination generates complex multi-annual wave patterns, with high-incidence cycles driven by seasonality and low-incidence intervals driven by behavioral feedback. By mapping stability as a function of seasonal forcing, mitigation strength, and cooperativity, we identify distinct period-doubling cascades with chaotic signatures arising from different mechanisms: the interplay between seasonality and behavior, and inter-pathogen cooperativity with the backward bifurcation it induces. We then analyze how these mechanisms interact across parameter ranges. Altogether, we show that cooperation fundamentally expands the spectrum of possible epidemic patterns, highlighting the importance of considering multi-pathogen interactions in epidemic modeling and control strategies.
\end{abstract}

\clearpage
\section{Introduction}

Emergent and re-emergent infectious diseases can severely disrupt societal function and well-being. Whether an outbreak permeates large fractions of the population depends on the interplay between pathogen and host population, modulated by internal and external forces. Key mechanisms include societal self-regulation via information feedback loops \cite{d2022behavioral,wagner2025societal,manfredi2013modeling,d2009information,donges2022interplay} and seasonality \cite{fisman2007seasonality,pascual2005seasonal,altizer2006seasonality}. These factors induce complex dynamics in both epidemic and endemic states \cite{stollenwerk2022seasonally,contreras2023emergency,siettos2013mathematical}.

However, diseases rarely occur in isolation: pathogens spread simultaneously and interact through varying degrees of competition or cooperation \cite{shaw2025co}. While competition, driven by mechanisms such as cross-immunity, active mitigation, or depletion of the susceptible pool, is well documented \cite{rohani2003ecological,contreras2023model,orostica2024early}, pathogen cooperation presents distinct modeling challenges. In this regime, one pathogen facilitates co-infection or subsequent infection with another, often by compromising physiological barriers or suppressing immune responses. For instance, some influenza viruses damage the respiratory epithelium, increasing susceptibility to secondary bacterial infections \cite{Cooperationbtw_Infl_StrepPneu,mccullers2014copathogenesis,morris2017secondary}. Similar facilitation is observed in HIV and hepatitis C co-infections \cite{Cooperationbtw_HIV_HCV_A,Cooperationbtw_HIV_HCV_B,Cooperationbtw_HIV_HCV_C}, among human papillomavirus (HPV) types \cite{Cooperationbtw_PapillomaVirusTypes}, in bacterial co-infections associated with SARS-CoV-2 \cite{CoInfectionNumbers_SARS-CoV-2}, and in the mutual enhancement between influenza and invasive pneumococcal disease \cite{CoInfectionNumbers_SARS-CoV-2_mortality,Cooperationbtw_Infl_IPD,klugman2009deadly}. However, the coupling between infections and interventions can also be behavioral: due to risk compensation \cite{messiah2012risk}, interventions against HIV can also facilitate the spread of other sexually transmitted infections \cite{mueller2025paradox, mallick2026stability}.

Mechanistically, multipathogen interaction has been modeled at varying complexity, from sequential coupling \cite{steindorf2022backward, aguiar2024backward} to systems incorporating full interactions and all possible disease states \cite{kramer2025limitations,PaperCoopSyndemics2013}. The competition-cooperation spectrum can be abstracted into a single coupling parameter $C$, which modulates the force of infection during co-infection. Here, $C > 1$ indicates cooperation, while $C < 1$ implies competition \cite{PaperCoopSyndemics2013}. In a symmetric susceptible-infectious-removed (SIR), SIR-SIR setting, \cite{PaperCoopSyndemics2013,CoopSyndemics_Khazaee_2022,ExactSXP_Paper} identified an "explosive" acceleration of contagion when $C\geq 2$. Even for overall reproduction numbers below one, this interaction drives a phenomenon resembling a first-order phase transition, explained by a positive feedback loop between existing and new infections. In models with waning immunity, such transitions often arise from the system tipping across the boundaries of attractor basins, typically via a backward bifurcation. Mechanisms known to induce such bifurcations include exogenous reinfection (e.g., tuberculosis) \cite{BckwBif_DueToExReinf3,BckwBif_DueToExReinf1,BckwBif_DueToExReinf2}, superinfection \cite{BckwBif_DueToSuperInfPapilloma}, relapse \cite{BckwBif_DueToRelapse}, imperfect vaccination \cite{BckwBif_DueToVacc1,BckwBif_DueToVacc2,BckwBif_DueToVacc3}, population heterogeneity \cite{BackwardsBif_CausesandExamples}, and antibody-dependent enhancement (as in dengue virus) \cite{steindorf2022backward, aguiar2024backward}.

In this paper, building on the multi-pathogen SIR-SIR configuration of \cite{PaperCoopSyndemics2013,CoopSyndemics_Khazaee_2022,ExactSXP_Paper}, we investigate the interplay between waning immunity, behavioral feedback, and seasonality in a SIRS coinfection model, proceeding in three stages. First, we examine the SIRS-SIRS model without external forcing and show that the first-order phase transition persists and is amplified: for reproduction numbers below one, a bistable region emerges in which a backward bifurcation drives the transition between the disease-free equilibrium (DFE) and the endemic equilibrium (EE). Second, introducing mitigation through behavioral feedback and seasonality individually, we find that each can generate limit cycles near the tipping point. Finally, we map the stability of the full system as a function of seasonal forcing and mitigation strength, revealing period-doubling cascades to chaos and alternating wave regimes around the critical reproduction number: one driven by seasonality (high incidence, fast dynamics) and the other by mitigation (low incidence, slow dynamics).

\section{Methods}
\subsection{Model backbone}\label{sec: the system}

\begin{figure}[H]
    \includegraphics[width=\linewidth]{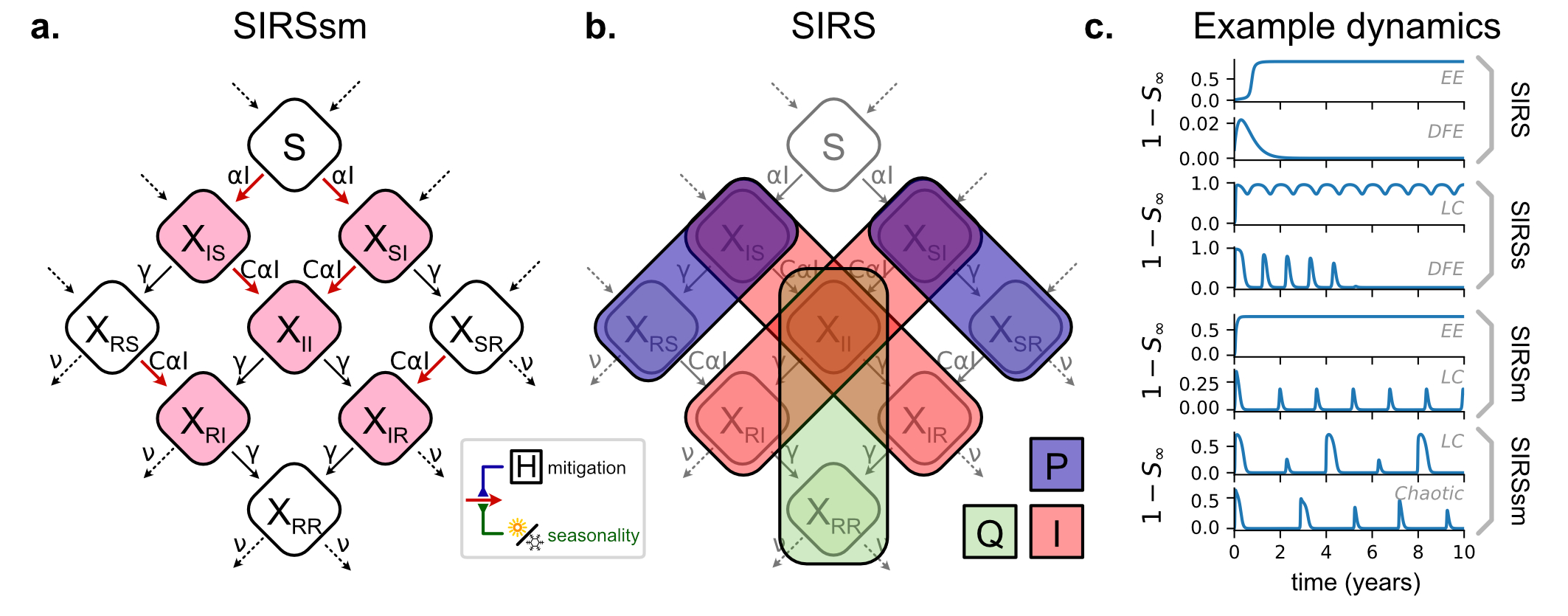}
    \caption{\textbf{Model overview. a.} SIRS coinfection model including seasonality and societal self-regulation through a hazard-mediated behavioral feedback (SIRSsm), representing equations~\eqref{eq:SIRSsm_first}--\eqref{eq:SIRSsm_last}. Infected compartments are marked in pink; their sum is the total number of infected individuals $I_{\rm tot}$. \textbf{b.} SIRS coinfection model without seasonality or behavioral feedback and its reduced variables (SIRS), representing equations~\eqref{eq:SIRS_first}--\eqref{eq:SIRS_last}. In our analysis, we assume symmetric conditions in initial conditions and transition parameters, so that $X_{xy} = X_{yx}$ at all times. Note that, for both models, immunity is lost, so individuals transition from the corresponding $R$ to $S$ compartments, as if the diagram were drawn on a torus. In panel \textbf{b}, the reduced model groups compartments into $P=X_{IS}+X_{RS}$ (individuals who have had contact with exactly one disease) and $Q=X_{II}+X_{RR}$ (individuals in the same state for both diseases); these groupings and all symbols are listed in \autoref{tab: Overview Variables}. \textbf{c.} Example long-term dynamics observed in each model variant (SIRS, SIRSs, SIRSm, SIRSsm). EE: endemic equilibrium; DFE: disease-free equilibrium; LC: limit cycle.}
    \label{fig:Figure_1}
\end{figure}

We use a compartmental model to describe the simultaneous spread of two pathogens within a single population, accounting for waning immunity and mechanisms of interaction among pathogens (competition-cooperation~\cite{PaperCoopSyndemics2013,CoopSyndemics_Khazaee_2022}), society (behavioral feedback~\cite{d2022behavioral,donges2022interplay}), and the environment (seasonality~\cite{wagner2025societal}). The model follows a susceptible-infectious-recovered-susceptible (SIRS) architecture with full interaction between the two pathogens, yielding 9 compartments (see~\autoref{fig:Figure_1}). All model variables and parameters are summarized in \autoref{tab: Overview Variables} and \autoref{tab: Overview Parameters}.

\begin{table*}[ht]
\centering
\makebox[\textwidth][c]{
\begin{tabular}{llp{7.5cm}}\toprule
\textbf{Variable} & \textbf{Definition} & \textbf{Interpretation} \\\midrule
$X_{xy}$ &- & Pop.\ fraction in disease state $(x,y)\in[S,I,R]^2$, symmetric under exchange of the two diseases ($X_{xy}=X_{yx}$). \\
$S$ & $X_{SS}$ & Pop.\ fraction susceptible to both diseases \\
$P$ & $\A+\aVar$ & Pop.\ fraction that has had contact with exactly one disease (susceptible to the other); grouping used in the reduced model (\autoref{fig:Figure_1}b)\\
$Q$ & $\AB+\ab$ & Pop.\ fraction in the same state for both diseases (infected with both or recovered from both); grouping used in the reduced model \\
$\X$ & $\A+\AB+\Ab$ & Pop.\ fraction infectious with one disease \\
$I_{\rm tot}$ & $\A+\B+\AB+\Ab+\aB$ & Total infected population fraction \\
$H$ & \autoref{eq: Res: Hazard} & Risk awareness in the population\\
\bottomrule
\end{tabular}}
\caption{\textbf{Model variables.} For each variable $X$ listed in the table, $X(0)$ denotes its initial value, $X_{\infty}$ the numerically obtained long--term behavior in the full model, and $X^{*}$ the analytical fixed point of the reduced model with time-independent spreading rates.}
\label{tab: Overview Variables}
\end{table*}

\begin{table*}[ht]
\makebox[\textwidth][c]{
\begin{tabular}{llp{5.5cm}ll}\toprule
\textbf{Category} & \textbf{Parameter} & \textbf{Interpretation} & \textbf{Baseline value} & \textbf{Role in this study}\\\midrule
\multirow{1}{*}{Initial Conditions} & $\epsilon$ & Fraction of initially infected & 0.5\% & Varied ($0.2$--$3\%$)\\
\midrule
\multirow{5}{*}{Disease Rates}
    & $\gamma_0$& Reference recovery rate & 0.1 $\text{days}^{-1}$ & Fixed\\
    & $\nu_0$& Reference waning immunity rate & 0.01 $\text{days}^{-1}$ & Fixed\\
    & $C$ & Cooperation value & 15 & Varied ($2$--$25$)\\
    & $R_0$ & Basic reproduction number & $0.9,~1.2,~1.2$* & Set per model\\
    & $\kappa$ & Rate-scaling parameter (disease time-scale) & 1 & Varied ($0.2$--$1$)\\
\midrule
\multirow{2}{*}{Seasonality}
    & $\omega$ & Seasonality frequency (seasonal time-scale) & $\frac{2\pi}{360}$ $\text{days}^{-1}$ & Fixed\\
    & $s$ & Seasonality amplitude & $0.25,~0,~0.25$* & Varied ($0$--$0.5$)\\
\midrule
\multirow{4}{*}{Behavioral Feedback}
    & $H_{\rm thresh}$ & Characteristic risk awareness & $\frac{2}{1000}$ & Fixed\\
    & $\epsilon_m$ & Softplus smoothness parameter & $\frac{1}{2000}$ & Fixed\\
    & $\tau_m$ & Risk perception delay (behavior time-scale) & 36 $\text{days}$ & Fixed\\
    & $m_{\rm max}$ & Maximum mitigation value & $0,~0.2,~0.2$* & Varied ($0$--$0.75$)\\
\bottomrule
\end{tabular}}
\caption{\textbf{Model parameters.} The final column indicates whether each parameter is held fixed throughout or systematically varied to generate the bifurcation diagrams (explored range in parentheses); baseline values are those used when the parameter is held fixed. Parameters marked with * take different baseline values depending on the included extension (seasonality $+s$, mitigation $+m$, or both $+sm$), listed in that order. For $R_0$, the value is set per model so that the effective reproduction number $R_t$ can explore both sides of the bifurcation, enabling tipping between attractors.}
\label{tab: Overview Parameters}
\end{table*}

The system of ordinary differential equations for the nine population fractions is:

\begin{align}
\label{eq:SIRSsm_first}
    \dot S&=-2\alpha \X\cdot S+\nu (\aVar+\bb)\\
\label{eq:SIRSsm_first infected}
    \dot \A&=+\alpha \X\cdot S-C \alpha \X\cdot \A-\gamma \A+\nu \Ab\\
    \dot \B&=+\alpha \X\cdot S-C \alpha \X\cdot \B-\gamma \B+\nu \aB\\
    \dot {\AB}&=+C \alpha \X(\A+\B)-2\gamma \AB\\
    \dot{\Ab}&=+C \alpha \X\cdot \bb+\gamma(\AB-\Ab)-\nu \Ab\\
\label{eq:SIRSsm_last infected}
    \dot{\aB}&=+C \alpha \X\cdot \aVar+\gamma(\AB-\aB)-\nu \aB\\
    \dot \aVar&=+\gamma \A-C \alpha \X\cdot \aVar-\nu \aVar+\nu \ab\\
    \dot \bb&=+\gamma \B-C \alpha \X\cdot \bb-\nu \bb+\nu \ab\\
\label{eq:SIRSsm_last}
    \dot{\ab}&=+\gamma(\aB+\Ab)-2\nu \ab.
\end{align}

where $X_{xy}$ denotes the fraction of the population in disease state $x \in \{S, I, R\}$ with respect to disease A and state $y \in \{S, I, R\}$ with respect to disease B. Transitions between the $X_{xy}$ compartments are governed by the base spreading rate $\alpha$, the interaction parameter $C$, the recovery rate $\gamma$, and the immunity decay rate $\nu$. We also assume that facilitation persists after recovery: infection from $X_{RS}$ or $X_{SR}$ is still boosted by $C$ until immunity wanes. For clarity, we present the model directly in its symmetric form, using a single set of rates ($\alpha$, $\gamma$, $\nu$, $C$) shared by both diseases; the general, potentially asymmetric model is recovered by assigning disease-specific rates (e.g.\ $\alpha_A\neq\alpha_B$) and initial conditions.

The initial conditions are symmetric, so that the fraction of initially infected $\epsilon$ is evenly distributed on the infected compartments of both diseases

\begin{align}
    S(0) = 1 - \epsilon \text{ and } \A(0) = \B(0) = \frac{\epsilon}{2}.
\end{align}

Variables $\XA=\A+\AB+\Ab$ and $\XB=\B+\AB+\aB$ represent the total fractions of the population infected with Disease A or B. Due to the symmetry of both the ODE system and the initial conditions under exchange of the two diseases A and B, all variables that are symmetric with respect to the disease type remain equal, $X_{xy}=X_{yx}$, which implies $\XA = \XB =: \X$. Due to the fixed total population, we have

\begin{align}
\label{eq: total pop=1, SIRS, original Variables}
\begin{split}
    1&=S+\A+\B+\AB+\aVar+\bb+\Ab+\aB+\ab\\
    &\stackrel{\rm symm.}{=} S+2\A+\AB+2\aVar+2\Ab+\ab.
\end{split}
\end{align}

The basic reproduction number of the system is the ratio of the spreading and recovery rates, as calculated in \autoref{sec: appendix R0 calc} through the Next Generation Matrix method \cite{NextGenMatrix,OnR0Def}. We define the basic recovery rate $\gamma_0$, the basic waning immunity rate $\nu_0$, the basic reproduction number $R_0$, and the rate-scaling parameter $\kappa$, which controls the relative speed of the disease dynamics with respect to the seasonal forcing (a larger $\kappa$ slows all disease transitions relative to the fixed seasonal period). These parameters determine the infection, recovery, and waning immunity rates as

\begin{equation}
\label{eq: total rates}
    \alpha_t = \frac{\gamma_0}{\kappa}\,R_0\Gamma(t)(1-m(H)),\quad\gamma = \frac{\gamma_0}{\kappa},\quad \nu    = \frac{\nu_0}{\kappa},
\end{equation}

yielding an effective reproduction number of

\begin{align}
\label{eq: R(t,h)}
    R_t=\frac{\alpha_t}{\gamma}=R_0\Gamma(t)(1-m(H)),
\end{align}

where $\Gamma(t)$ and $(1 - m(H))$ represent the seasonal forcing and the behavioral feedback, respectively. Here $m(H)\in[0,m_{\rm max}]$ is the mitigation intensity, an increasing function of the population's risk awareness $H$ that is specified in \autoref{eq: Res: mitigation(h)} below, and $\alpha_t$ denotes the resulting time-dependent effective spreading rate. We assume a sinusoidal form for the seasonal forcing:

\begin{align}
    \Gamma(t)=(1+s\cos(\omega t)),
\end{align}

where $s$ is the amplitude of seasonality and $\omega$ is the seasonal frequency. The behavioral feedback and mitigation component is described in the following section.

\subsection{Behavioral feedback}

Following \cite{d2022behavioral,InclBeh_Onofrio_Old_Vacc_ErlangKernel,wagner2025societal}, we define a hazard variable $H$ quantifying risk perception in the population. We assume that $H$ depends on the history of the total number of infected individuals, $\X_{\rm tot} = \A + \Ab + \AB + \aB + \B$ (sum of the compartments marked in red in \autoref{fig:Figure_1}a), so that:

\begin{align}
\label{eq: Res: Hazard}
    H(t)&=\int_{-\infty}^t\X_{\rm tot}(s)K(t-s)ds,\\
\label{eq: Res: Erlang Kernel (2)}
    K(t)&=\frac{t}{\tau_m^2}e^{-\frac{t}{\tau_m}}.
\end{align}

For computational efficiency, as done in \cite{d2009information,wagner2025societal}, we use the properties of the Erlang kernel to replace the integro-differential equation by two auxiliary ODEs:

\begin{align}
\dot H(t) &= \frac{1}{\tau_m}\left(H'(t)-H(t)\right),\\
\dot H'(t) &= \frac{1}{\tau_m}\left(I_{\rm tot}(t)-H'(t)\right),
\end{align}

where $\tau_m$ is the time it takes for information about the number of infected to affect the Hazard maximally. The mitigation implemented in response to $H$ is given by:
\begin{align}
\label{eq: Res: mitigation(h)}
    m(H)&=m_{\rm max}-\frac{m_{\rm max}}{H_{\rm thres}}\epsilon_m \log\left(1+\exp\left(\frac{H_{\rm thres}-H}{\epsilon_m}\right)\right)\in[0,m_{\rm max}],
\end{align}

where $m_{\rm max}$ represents the maximum mitigation value, and $H_{\rm thres}$ the reaction threshold where mitigation saturates \cite{donges2022interplay}. The effective mitigation modulates the spreading rates and the reproduction number by a factor of $(1-m(H))$.

Depending on whether we include seasonality or behavioral feedback, we refer to the model as SIRSs (seasonality only), SIRSm (mitigation only), or SIRSsm (both). Representative long-term dynamics for each variant are shown in \autoref{fig:Figure_1}c. The reference value $R_0$ is chosen so that the system dynamics remain centered in a physically meaningful regime, i.e., one with more attractors than just a stable disease-free equilibrium.

\section{Analytical study of the SIRS-SIRS model}
\subsection{Equilibria and bifurcations}\label{sec: analytical Results}

We first determine the long-term behavior of the SIRS-SIRS model in the absence of seasonality and behavioral feedback. However, in contrast with the SIR-SIR case, the additional waning-immunity fluxes prevent several groupings of variables. For instance, $\B$ and $\bb$ cannot be combined because the waning flux $\nu \bb$ returns to $S$ only from $\bb$ but not from $\B$. Nevertheless, the symmetry in the disease type $X_{xy} = X_{yx}$ still holds. We thus introduce $P := \A + \aVar = \B + \bb$, which represents individuals who have had or currently have contact with exactly one of the two diseases (as in \cite{PaperCoopSyndemics2013}), and $Q :=\AB + \ab$, defined for practical purposes. With these identifications, the system reduces to the variable set $\mathcal X := (S,\, P,\, \X,\, Q,\, \A,\, \AB)^T$, illustrated in \autoref{fig:Figure_1}b. The flux $\nu \bb$ can now be written as $\nu (P - \B) = \nu (P - \A)$ using only variables from the reduced set. The reduced ODE system now reads:

\begin{align}
\label{eq:SIRS_first}
    \dot S&=-2\alpha \X\cdot S+2\nu (P-\A),\\
\label{eq:SIRS_dot P}
    \dot P &=+\alpha \X\cdot S-C \alpha \X\cdot P+\nu(\X-P+Q-2\AB),\\
\label{eq:SIRS_dot X}
    \dot \X&=+\alpha \X\cdot S+C \alpha \X\cdot P-\gamma \X,\\
\label{eq:SIRS_dot Q}
    \dot Q&=2C \alpha \X \A+2\gamma(-2\AB+\X-\A)-2\nu(Q-\AB),\\
\label{eq:SIRS_dot A}
    \dot \A&=\alpha \X S-C \alpha \X \A-\gamma \A+\nu(\X-\AB-\A),\\
\label{eq:SIRS_last}
    \dot{\AB}&=2C \alpha \X \A-2\gamma \AB.
\end{align}

The assumption of demographic equilibrium in this reduced set of variables becomes
\begin{align}
\label{eq: total pop=1, SIRS, SXPQAAB-Var}
    1\stackrel{\eqref{eq: total pop=1, SIRS, original Variables}}{=}S+\underbrace{2\A+\AB+2\Ab}_{2\X-\AB}+\underbrace{2\aVar}_{2P-2\A}+\underbrace{\ab}_{Q-\AB}=S+2\X-2\AB+2P-2\A+Q.
\end{align}

Similarly, the initial condition now reads:

\begin{align}
\label{eq: Res: SPXU InCond}
    \A(0)=\B(0):=\frac{\epsilon}{2} \text{ and } S(0)&:=1-\epsilon \\
    \Rightarrow \mathcal X(0)=(S(0),~P(0),~\X(0),~Q(0),\A(0),\AB(0))&=\left(1-\epsilon,~\frac{\epsilon}{2},~\frac{\epsilon}{2},~0,~\frac{\epsilon}{2},~0\right).
\end{align}

The fixed points of the reduced model satisfy $\dot{\mathcal X}=0$ (see~\autoref{sec: appendix expl FP calc} for details). The trivial solution corresponds to the system's disease-free equilibrium (DFE):

\begin{align}
\label{eq: DFE}
(S^*,\X^*,P^*,Q^*,\A^*,\AB^*)=(1,0,0,0,0,0).
\end{align}

Regarding non-trivial solutions, we obtain

\begin{align}
\label{eq: S(P,X)}
    S^*&=\frac{1}{\alpha}(\gamma-C \alpha P^*)\\
\label{eq: A(P,X)}
    \A^*&=P^*+\frac{1}{\nu}(C \alpha P^*-\gamma)\X^*\\
\label{eq: AB(P,X)}
    \AB^*&=\frac{C \alpha}{\gamma}\X^* P^*+\frac{C \alpha}{\gamma\nu}(C \alpha P^*-\gamma){\X^*}^2\\
\label{eq: Q(P,X)}
    Q^*&=P^*-\rho \X^*+2\frac{C \alpha\rho}{\gamma}\X^* P^*+\frac{C \alpha}{\gamma\nu}(C \alpha P^*-\gamma){\X^*}^2,\\
\label{eq: P(X)}
    P^*&=\frac{C \alpha\gamma\rho {\X^*}^2+\nu\gamma\rho^2 \X^*}{C \alpha^2\rho {\X^*}^2+C \alpha(\gamma+\nu\rho^2)\X^*+\nu\gamma\rho},
\end{align}
where $\rho:=1+\frac{\gamma}{\nu}$ and $\X^*$ is the solution of a cubic polynomial

\begin{align}
\label{eq: general cubic X polynomial}
0=f(\X^*)=a{\X^*}^3+b{\X^*}^2+c\X^*+d
\end{align}

with the pre-factors

\begin{align}
\label{eq: Prefac of CbEq for X}
\begin{split}
    a &= \left(C\alpha\right)^2\rho^2>0,\\
    b &= \rho C \alpha(2\gamma+\nu\rho^2-C \alpha),\\
    c &= \left[C \alpha\left(\frac{\gamma^2}{\alpha}-\gamma-\nu\rho^2\right)+2\nu\gamma\rho^2\right],\\
    d &= \left(\frac{\gamma}{\alpha}-1\right)\nu\gamma\rho.
\end{split}
\end{align}

\begin{figure}[H]
    \centering
\includegraphics[width=150mm]{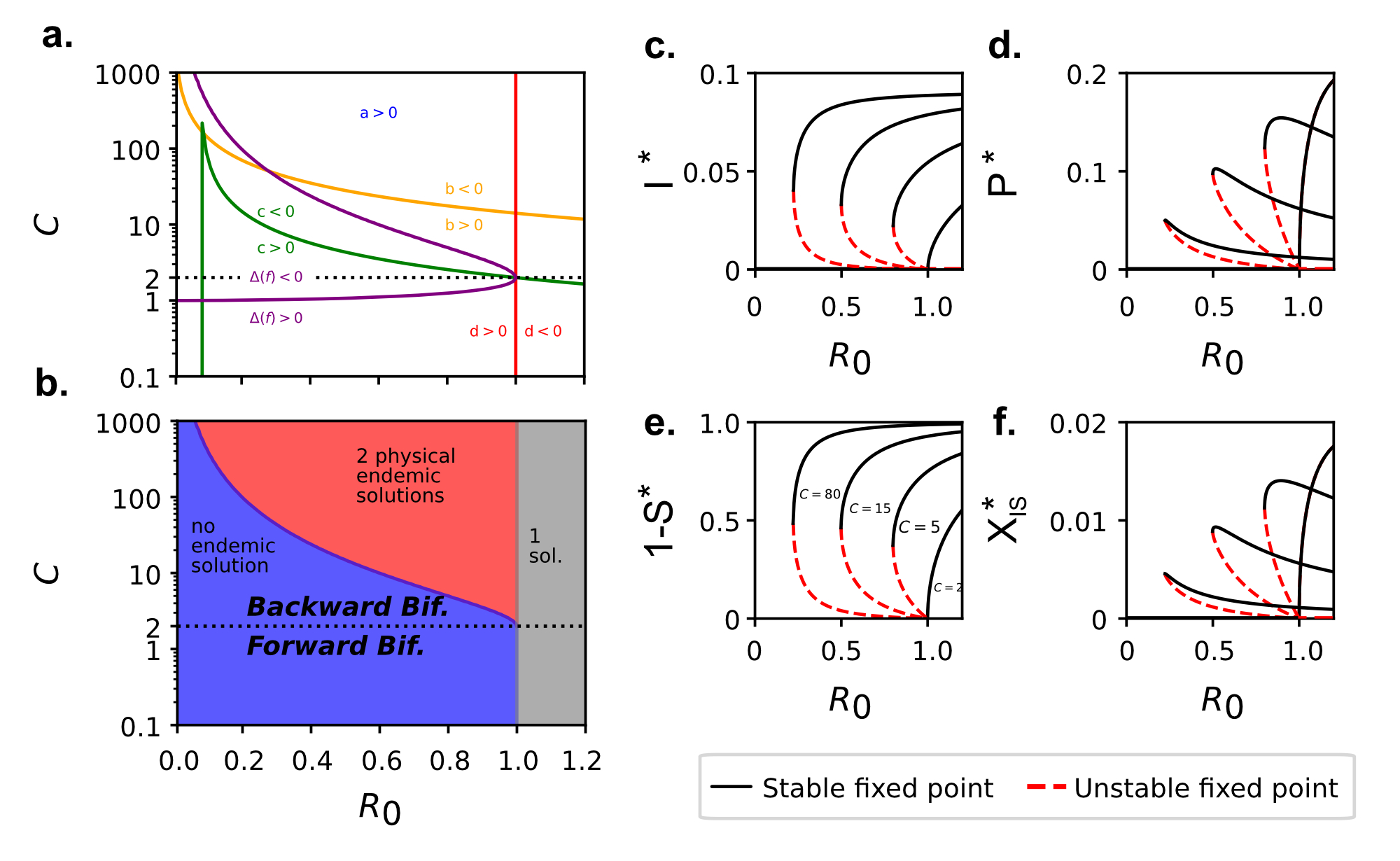}\caption{\textbf{a.} Signs of the cubic equation prefactors and discriminant. \textbf{b.} Parameter regions colored by the number of physically meaningful endemic solutions, i.e.\ positive branches of the endemic equilibrium (the disease-free and negative solutions are not counted); a black horizontal line separates regions of forward and backward bifurcations. \textbf{c--f.} Projection of multidimensional fixed points and their stability (solid line represents stable, dashed lines represent unstable), for different values of $C$.}
\label{fig:Figure_2}
\end{figure}

By the rule of discriminants, our cubic has three distinct real solutions if and only if the discriminant of $f$
\begin{align}
    \Delta(f)[a,b,c,d]\equiv 18abcd - 4b^3d + b^2c^2 -4ac^3 - 27a^2d^2
\end{align}
is positive. Two of these solutions correspond to the two branches of the endemic equilibrium, while the third is always negative and thus unphysical. When the discriminant is negative, only this negative solution is real. At $\Delta(f)=0$, $f$ has a double root: the bifurcation point at which the endemic equilibrium emerges. If this happens at positive values of the phase space variables, we have two physical branches of the endemic equilibrium forming a backward bifurcation \cite{BackwardsBif_CausesandExamples}. \autoref{fig:Figure_2}a shows the sign of the discriminant and the curve along which it vanishes, where the bifurcation occurs. For $C=2$, the $R_0$ value of the bifurcation point is at its maximum $R_0=1$.

At first glance, the bifurcation point seems to shift to lower $R_0$ values as $C$ increases from $2$. However, only the bifurcation occurring at $C>2$ is physically meaningful, since the one at $C \in [1,2]$ corresponds to the emergence of endemic branches at negative $\X^*$ values. To verify this, we apply Descartes' Rule of Signs: the maximum number of positive roots of a polynomial $P_n(x) = \sum_{k=0}^n a_kx^k$ equals the number of sign changes in the ordered coefficient set $\{a_k\}_{k=0}^n$ \cite{DescartesRuleOfSigns_OldPaper,DescartesRuleOfSigns_Proof}, and the maximum number of negative roots equals the number of sign changes in the set $\{(-1)^k a_k\}_{k=0}^n$. Fewer positive or negative fixed points may occur if some roots are complex, i.e., if the discriminant is negative. In our case, the only physical solutions are the branches of the endemic equilibrium, which only become real for $\Delta(f)>0$. Therefore, examining the signs of the coefficients \eqref{eq: Prefac of CbEq for X} within the region of positive discriminant (see \autoref{fig:Figure_2}a) determines the number of physically relevant solutions for $\X^*$. As shown in \autoref{fig:Figure_2}b, for $C<2$ there are no positive solutions until $R_0$ exceeds 1, at which point one endemic equilibrium branch emerges above 0. In contrast, for $C>2$, both endemic equilibrium branches emerge through a bifurcation where the discriminant vanishes. For $R_0>1$, the lower branch becomes negative, leaving only the upper branch as a physically meaningful steady state.

Stability of the fixed points follows from the eigenvalues of the reduced system's Jacobian matrix evaluated at the corresponding equilibria,

\begin{align}
\label{eq: Jacobian}
    J\stackrel{\eqref{eq:SIRS_first}-\eqref{eq:SIRS_last}}{=}\begin{pmatrix}
        -2\alpha \X & 2\nu             & -2\alpha S                  &  0  & -2\nu                 & 0\\
        \alpha \X   & -C \alpha \X-\nu & \alpha S-C \alpha P+\nu     & \nu &    0                  & -2\nu\\
        \alpha \X   & C \alpha \X      & \alpha S+C \alpha P-\gamma  &  0  &    0                  & 0\\
        0           & 0                & 2C \alpha \A+2\gamma        &-2\nu&  2C \alpha \X-2\gamma & -4\gamma+2\nu\\
        \alpha \X   & 0                & \alpha S-C \alpha \A+\nu    &  0  &-C \alpha \X-\gamma-\nu& -\nu\\
        0           & 0                & 2C \alpha \A                &  0  & 2C \alpha \X          & -2\gamma
    \end{pmatrix}.
\end{align}

Due to the normalization constraint \eqref{eq: total pop=1, SIRS, SXPQAAB-Var}, the system's equations are linearly dependent, yielding a first eigenvalue of $0$. At the disease-free equilibrium \eqref{eq: DFE}, the Jacobian reads

\begin{align}
\label{eq: DFE Jacobian}
    J_{\rm DFE}\stackrel{\eqref{eq: Jacobian}\eqref{eq: DFE}}{=}\begin{pmatrix}
        0 & 2\nu & -2\alpha      &  0  & -2\nu     & 0\\
        0 & -\nu & \alpha+\nu    & \nu &    0      & -2\nu\\
        0 & 0    & \alpha-\gamma &  0  &    0      & 0\\
        0 & 0    & 2\gamma       &-2\nu& -2\gamma  & -4\gamma+2\nu\\
        0 & 0    & \alpha+\nu    &  0  &-\gamma-\nu& -\nu\\
        0 & 0    & 0             &  0  & 0         & -2\gamma
    \end{pmatrix},
\end{align}
and the non-zero eigenvalues are $-\nu,\alpha-\gamma,-2\nu,-\gamma-\nu,-2\gamma$. Consequently, the DFE is stable iff $\alpha<\gamma$, i.e., $R_0<1$. For the endemic branches, we compute the eigenvalues numerically. The upper branch is stable and the lower branch unstable, consistent with the DFE's stability behavior.

We obtain a backward bifurcation for $C>2$ and a forward bifurcation for $C\leq2$, qualitatively matching the result of the lower dimensional model from \cite{Chen2017Fundamental}. The resulting physical equilibria are shown in \autoref{fig:Figure_2}c--f for different phase-space variables at different $C$ values. Note that the figure shows one-dimensional projections of higher-dimensional fixed points. Hence, the projection for the unstable fixed-point does not correspond to the exact boundary between basins of attraction.

The backward bifurcation we observe arises from the additional compartment $P$, whose members acquire the second disease at the cooperativity-boosted rate $C \alpha \X$: this flux toward (co-)infection is stronger than the rate $\alpha \X$ at which fully susceptible individuals in $S$ acquire their first infection. When the susceptible population $S$ becomes infected, it decreases, reducing the flux $\alpha S\X$ into the infected compartment; this mechanism typically generates a forward bifurcation. However, in our setting, half of the individuals leaving $S$ transition to $P$, where their contribution to the infection flux increases to $C \alpha P\X$. The total contribution to the infected compartment is thus $\dot \X\propto [\alpha S+C \alpha P-\gamma]\propto[R_0S+CR_0P-1]$. If this additional influx from $P$ accounts for the loss from $S$, then the endemic and the disease-free equilibrium can coexist at $R_t<1$, and a backward bifurcation occurs. Since $P$ increases by half the amount that $S$ decreases, the criterion for a backward bifurcation, $0 < \Delta \dot{\X} \propto [R_0 \Delta S + C R_0 \Delta P] \propto [-2R_0 + C R_0]$, is only fulfilled if individuals joining $P$ at least double their own flux toward $\X$ ($C \geq 2$), matching the results from the model with persistent immunity \cite{PaperCoopSyndemics2013}.

\subsection{Backward bifurcation-induced first-order phase transition}
\label{sec: myResults: numerical}

\begin{figure}[H]
    \centering\includegraphics[width=88mm]{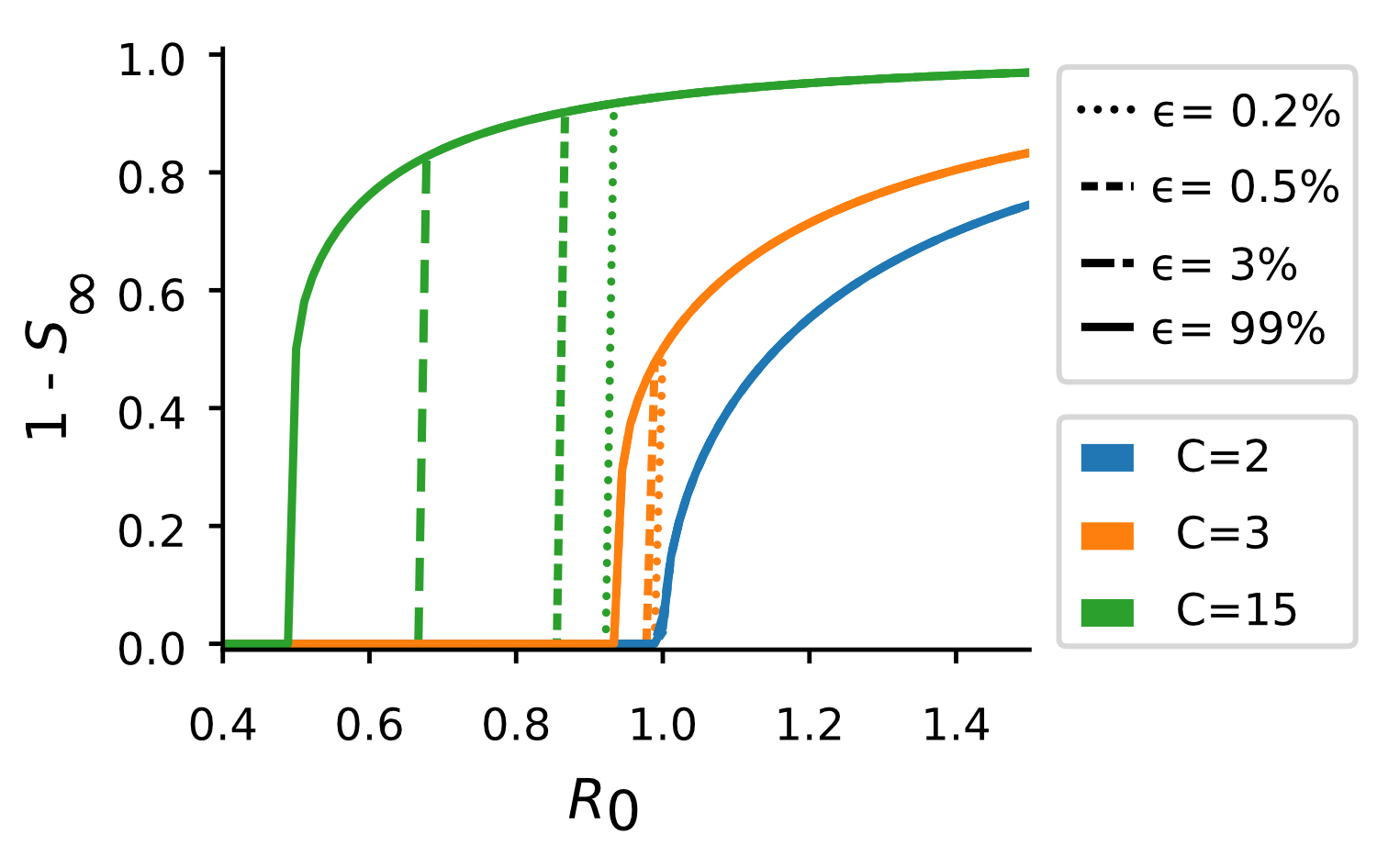}\caption{\textbf{Backward-bifurcation-induced first-order phase transition of the SIRS--SIRS model.} Depending on the cooperation strength $C$, cooperative symmetric SIRS--SIRS systems may exhibit a first-order phase transition in the asymptotic prevalence, $1-S_\infty$, as trajectories shift from the disease-free to the endemic basin of attraction. The critical reproduction number for this transition also depends on the initial infected fraction, $\epsilon$. Increasing $C$ lowers $R_0^{\rm crit}$ and raises $1-S_\infty$ for all $R_0$; increasing $\epsilon$ lowers $R_0^{\rm crit}$ without changing the endemic curve, with the $\epsilon$ dependence visible only near threshold, where the curves for $C=2$ and $C=3$ nearly overlap. As shown in Fig.~\ref{fig:Figure_2}b, the transition is continuous (forward bifurcation) for $C\leq2$ and discontinuous (backward bifurcation) for $C>2$.}
\label{fig:Figure_3}
\end{figure}

Just as in the case without waning immunity ($\nu=0$, see \cite{PaperCoopSyndemics2013}), the system exhibits a first-order phase transition in the order parameter $1-S_\infty$ and the control parameter $R_0$. The transition occurs at a value $R_0^{\rm crit}$ with bistability between the disease-free equilibrium and the stable endemic branch, due to the backward bifurcation discussed in \autoref{sec: analytical Results}. As shown in~\autoref{fig:Figure_3}, $R_0^{\rm crit}$ depends on $C$ and $\epsilon$: Increasing $C$ lowers $R_0^{\rm crit}$ and increases $1-S_\infty$ for all $R_0$ values, whereas increasing $\epsilon$ only lowers $R_0^{\rm crit}$, after which $1-S_\infty$ follows the curve set by the $C$ value. For sufficiently high $\epsilon$, further increases do not shift the critical value.

This behavior follows from the analytical equilibria derived in the previous section. For small $R_0$, only the DFE is stable. As $R_0$ increases, a region of bistability emerges, in which the DFE and a stable EE branch are connected by an unstable branch, located between their basins of attraction. The existence of this bistability region depends only on $C$, while $\epsilon$ determines when the border between the basins is crossed. The phase transition can therefore be interpreted as a tipping point over the basin border.

For $C> 2$, this is ensured to happen before the DFE loses stability at $R_0 = 1$. The reason for this is that, in the presence of a backward bifurcation, the endemic basin converges to the entire physically relevant phase space $\{R_\infty \mid R_\infty \in [0, \infty)\}$ as $R_0 \to 1^{-}$. The order of the transition is therefore set by the bifurcation type rather than by $\epsilon$. In particular, taking $\epsilon\to0$ does not recover a continuous transition when $C>2$: the critical value $R_0^{\rm crit}$ shifts toward $1^{-}$, but the jump in $1-S_\infty$ stays finite because the stable endemic branch remains bounded away from the DFE. A continuous phase transition is recovered only in the forward-bifurcation regime $C\leq2$.

\section{Numerical study of the SIRS-SIRS model with behavioral feedback and seasonality}
\subsection{SIRSs and SIRSm coinfection models}

The two extensions differ in the nature of the oscillations in the effective reproduction number $R_t$: seasonality induces them at a slow, constant rate, generating oscillations around the EE, whereas mitigation responds to the history of the phase-space variable $\X$ and can produce comparable oscillations in low-prevalence regimes.

\begin{figure}[H]
    \centering
        \includegraphics[width=130mm]{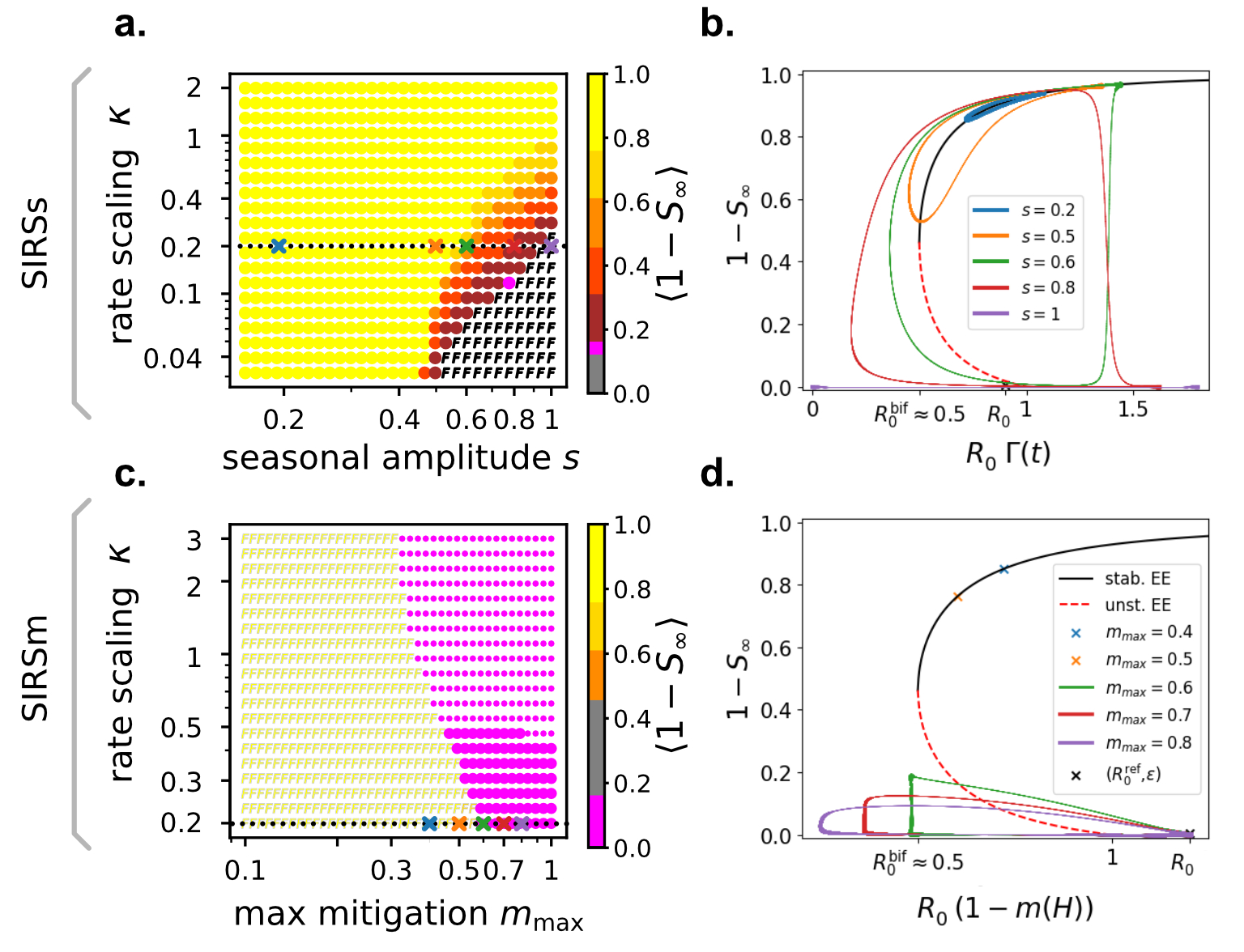}
    \caption{\textbf{Long-term dynamics of SIRSs (\textbf{a, b}) and SIRSm (\textbf{c, d}) coinfection models at $\epsilon=0.5\%$ and $C=15$}. \textbf{a.} Initializing the system at $R_0\approx0.9$, we show that the average prevalence (i.e., infected and recovered) $\langle1-S_\infty\rangle$ in the SIRSs coinfection model shows a discontinuous jump between DFE fixed points (F) and endemic limit cycles. Average prevalence is calculated over the last $30\%$ of a 25-year simulation period. \textbf{b} Long-term dynamics of sample trajectories at $\kappa=0.2$ and different seasonal amplitudes $s$. \textbf{c.} Initializing the system at $R_0=1.2$, we show that the SIRSm coinfection model exhibits either stable fixed points at high prevalence (F) or limit cycles at low prevalence ($\bullet$). Again, there is a sharp transition between both regions. As the period of attractors in the SIRSm model is not constrained to one year, unlike in the SIRSs case, the average prevalence is computed over the final $30\%$ of an extended 100-year simulation. \textbf{d.} Long-term dynamics of selected trajectories at $\kappa=0.2$ and different levels of maximum mitigation $m_{\rm max}$. The $\kappa$ axis in panels \textbf{a} and \textbf{c} is shown on a logarithmic scale because $\kappa$ enters the rates multiplicatively and spans a wide range of disease-to-forcing time-scale ratios, which resolves the fast- and slow-dynamics regimes evenly.}
    \label{fig:Figure_4}
\end{figure}

We find that both versions can exhibit limit cycles around the fixed points of the SIRS–SIRS model. The interesting dynamics arise in the bistability region between the endemic and disease-free equilibria. We set $\epsilon = 0.5\%$ and $C = 15$ and analyze the nature and approximate position of the attractors in phase space for both extensions, varying the forcing strength (seasonal amplitude $s$ or maximum mitigation $m_{\rm max}$) and the rate-scaling parameter $\kappa$, while fixing the time scale of the model extension (seasonality: $\omega$, behavioral feedback: $\tau_m$).

Given that the SIRSs model oscillates symmetrically around the reference $R_0$, we choose $R_0 = 0.9$, close to the tipping point at which the phase transition discussed in \autoref{sec: analytical Results} occurs. In~\autoref{fig:Figure_4}a, two regions with distinct long-term behavior separate clearly: trajectories converging to the DFE (stable fixed point, F) and high-prevalence limit cycles ($\bullet$). The disease-free region occurs at lower $\kappa$ and higher $s$, as expected: a larger seasonal amplitude allows $R_t$ to drop further below the tipping point, where the DFE basin of attraction occupies the entire phase space, and faster disease dynamics accelerate convergence toward the DFE (cf.~\autoref{fig:Figure_4}bc, purple line).

At the boundary between these two regions, the oscillation amplitude jumps, resembling a saddle-node on an invariant circle (SNIC) bifurcation. However, unlike SNIC bifurcations, the oscillation frequency does not increase continuously at the bifurcation point but remains fixed by the seasonal frequency $\omega$, as in a Hopf bifurcation. In the remaining parameter space, endemic limit cycles show a sharp transition between low prevalence (green and red in~\autoref{fig:Figure_4}b,c) and high prevalence (blue and yellow), depending on whether the trajectory temporarily crosses into the DFE basin of attraction in the SIRS–SIRS model (red dashed line in~\autoref{fig:Figure_4}b,c).

For oscillations induced by behavioral feedback, the oscillating factor $(1-m(H)) \in [0,1]$ is bounded from above by 1, so $R_0$ represents the maximum of $R_t$ rather than its mean. We therefore choose the higher value $R_0=1.2$, which allows $R_t$ to oscillate across the tipping point between the endemic and disease-free basins. Since $H(0)=0$, $R_t$ starts at this maximum, corresponding to a regime of increasing case numbers.

\begin{figure}
    \centering
    \includegraphics[width=88mm]{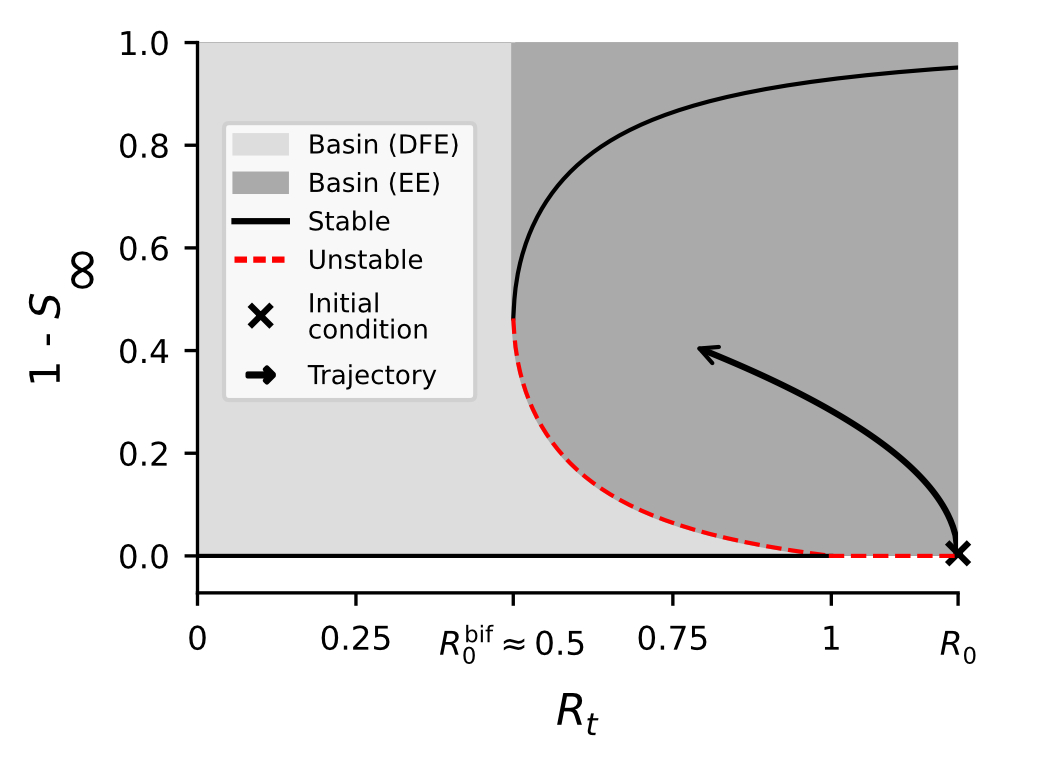}
    \caption{\textbf{Schematic representation of a behavior-driven trajectory in the projected phase space.} Starting at $R_0=1.2$, the behavioral feedback decreases $R_t$ as case numbers rise. Depending on $m_{\rm max}$ and the velocity of the disease dynamics, the trajectory may or may not tip onto the disease-free basin. The approximate positions of the basins of attraction are marked by the shaded backgrounds.
    }
    \label{fig:Figure_5}
\end{figure}

Simulations again reveal a clear separation of the parameter space into two regions (see \autoref{fig:Figure_4}c). Low $m_{\rm max}$ leads to endemic long-term behavior (in contrast to the SIRSs model, without sustained oscillations), whereas higher $m_{\rm max}$ drives the system towards the low-prevalence region, where both limit cycles and fixed points (DFE) are possible. Faster disease dynamics (low $\kappa$) expand the $m_{\rm max}$ interval over which the trajectory remains endemic, because the rapid vertical convergence in phase space primarily affects the endemic side of the bifurcation. Low $\kappa$ also lets cases rise further before mitigation acts, and higher $m_{\rm max}$ amplifies the deviations from equilibrium: both prolong the return to the original state, producing larger limit cycles in both phase and parameter space with correspondingly longer periods. The rapid response to rising cases arises from the steep initial slope of the memory kernel \eqref{eq: Res: Erlang Kernel (2)} and the assumed characteristic delay $\tau_m = 36$ days; the kernel’s long tail then sustains mitigation, so that convergence to the DFE appears as an almost vertical trajectory in phase space for some simulations (see~\autoref{fig:Figure_4}d). The resulting mitigation response is asymmetric: $R_0$ decreases rapidly when cases rise but increases only after cases have fully declined. This explains why limit cycles do not occur around endemic equilibria: $m_{\rm max}$ is insufficient to push the system into the disease-free basin. \autoref{fig:Figure_5} shows the projection of attractors and their basins of attraction.

\subsection{SIRSsm coinfection model}

We now analyze how seasonality and behavioral feedback jointly shape the long-term dynamics of the system. As in the SIRSm model, we fix $R_0=1.2$ to ensure the behavioral feedback oscillations span the relevant basins of attraction.

\begin{figure}
    \centering
    \includegraphics[width=0.9\linewidth]{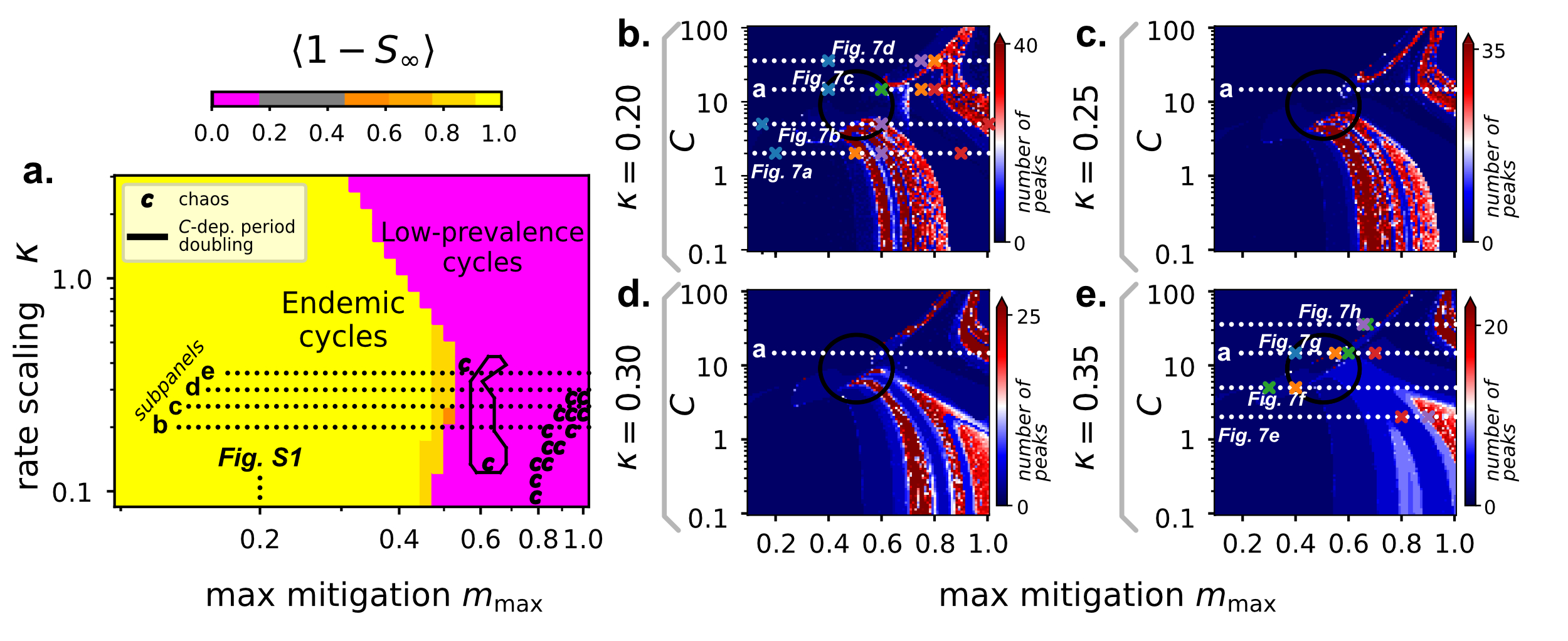}
    \caption{\textbf{Long-term dynamics of the SIRSsm coinfection model. a.} For $C=15$, dynamics are shown as a function of $\kappa$ and $m_{\rm max}$ ($s=0.25$). Across the parameter space, only limit cycles or more complex long-term behaviors are observed; fixed points are absent. In some regions, limit cycles bifurcate into multiple loops and transition to chaos. Color indicates the average prevalence over 100 years after a period of up to 900 years of transient. \textbf{b--e.} The number of peaks in the time series serves as a proxy for period-doubling cascades to chaos, identifying regions with chaotic signatures (red). Black circles mark approximate locations of cooperation-induced, non-chaotic period-doublings. The $\kappa$ axis is shown on a logarithmic scale, as in \autoref{fig:Figure_4}, because $\kappa$ scales all rates multiplicatively. Unspecified parameters use default values from \autoref{tab: Overview Parameters}.
    }
    \label{fig:Figure_6}
\end{figure}

\begin{figure}
    \centering
    \includegraphics[width=180mm]{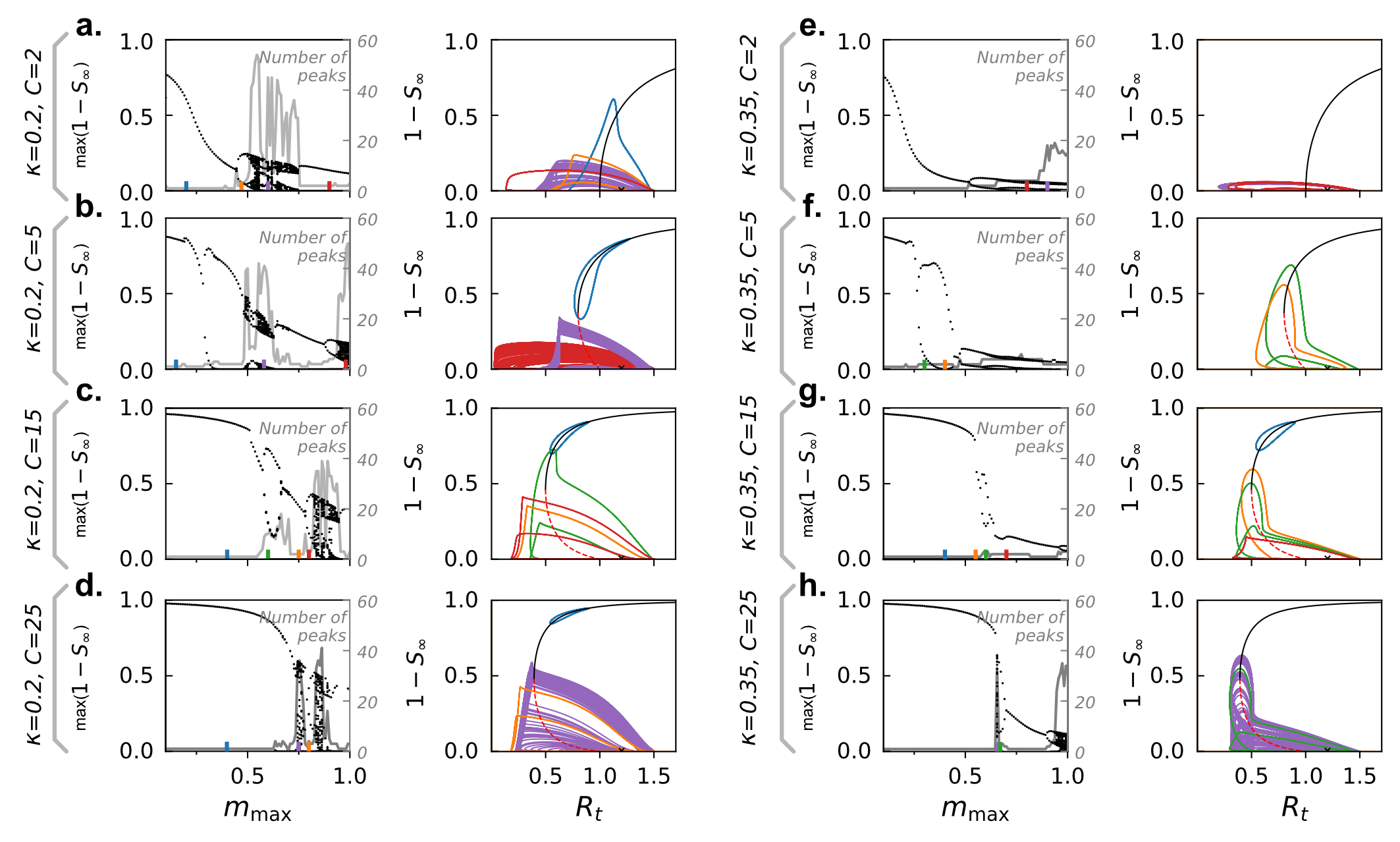}
    \caption{\textbf{Long-term behavioral changes as a function of maximum mitigation $m_{\rm max}$ for selected rate scaling $\kappa$ and cooperation $C$ at $s=0.25$. a--h.} The number of distinct peaks in the time series (in grey) and their $1-S_\infty$ value (in black dots) serves as an indicator of changes in solution behavior as $m_{\rm max}$ is varied, shown for $C=2,5,15,25$ at $\kappa = 0.2$ (\textbf{a--d}) and $\kappa = 0.35$ (\textbf{e--h}). As $m_{\rm max}$ increases, limit cycles undergo single period-doublings or cascades, with dynamics approaching the DFE. For $m_{\rm max}$ values where peaks drop close to the DFE, some may not be visible due to numerical precision. Right panels display representative trajectories for selected $m_{\rm max}$ values and the analytical backward bifurcation.
    }
    \label{fig:Figure_7}
\end{figure}

Depending on parameters, limit cycles occur at endemic or low-prevalence levels. When behavioral feedback predominates, the system shifts abruptly to low-prevalence cycles, consistent with the SIRSm coinfection model (\autoref{fig:Figure_6}a; see \autoref{fig:Figure_S1} for a similar analysis varying the seasonal forcing $s$ instead). As shown for single-pathogen systems~\cite{wagner2025societal}, the interplay between seasonality and feedback produces especially complex dynamics. With weak seasonality ($s = 0.25$), multidimensional period-doubling cascades emerge within the combined parameter space of $m_{\rm max}$, $C$, and $\kappa$ (see \autoref{fig:Figure_6} and \autoref{fig:Figure_7}). The number of distinct peaks in the time series serves as an early indicator of chaotic dynamics, delineating regions of regular and aperiodic behavior (\autoref{fig:Figure_6}b--e). In multipathogen systems, the interaction force facilitates cooperation-driven period-doublings and chaos, broadening the spectrum of dynamical regimes relative to single-pathogen models. Note that the influence of cooperation is not restricted to large $C$: as discussed below, the coexisting attractors that emerge at moderate cooperation first interrupt the period-doubling cascades driven by mitigation and seasonality, and chaotic signatures return as $C$ increases further (see \autoref{fig:Figure_6} and \autoref{fig:Figure_7}). We use large values ($C=15,~25$) to map the full extent of these regimes, not as empirical estimates (see Discussion).

Pathogen competition, i.e., the regime $C<1$, in which infection with one pathogen hinders acquisition of the other (as defined in the Introduction), intensifies the effects of mitigation, with regions displaying period-doubling cascades to chaos that closely parallel those observed in single-pathogen systems under analogous conditions~\cite{wagner2025societal,stollenwerk2022seasonally}. For $C > 2$, a backward bifurcation allows endemic and disease-free equilibria to coexist below $R_t=1$, producing endemic limit cycles at low seasonality and mitigation (see \autoref{fig:Figure_6}a). As cooperation increases, the endemic equilibrium diverges further from the DFE, requiring larger $m_{\rm max}$ values to restore low prevalence after the summer period (see \autoref{fig:Figure_7}). This increasing separation fragments the system's dynamics, interrupts period-doubling cascades, and temporarily suppresses chaotic behavior at intermediate cooperation ($C=5$ in \autoref{fig:Figure_7}; black circles in \autoref{fig:Figure_6}b--e). After several doublings ($C=5$ and $C=15$ in \autoref{fig:Figure_7}), chaotic signatures reappear ($C=25$), particularly in the left, cooperativity-driven segment of the Arnold tongue-shaped parameter region (see \autoref{fig:Figure_6}b--e). In contrast, the right tongue and regions with low $C$ reflect chaos primarily induced by mitigation. The cooperativity-induced cascade is confined to a limited $m_{\rm max}$ interval (black circles and left tongue in \autoref{fig:Figure_6}). When only mitigation is varied, these period-doublings are transient ($C=5$ and $C=15$, \autoref{fig:Figure_7}). For slower system dynamics (higher $\kappa$), the left tongue narrows further (see \autoref{fig:Figure_6}e). At $\kappa=0.35$, the cooperativity-induced cascade persists but shows chaotic signatures only within narrow $m_{\rm max}$ intervals (purple curve, \autoref{fig:Figure_7}h). Finally, most lower loops arising from period-doubling bifurcations converge toward the disease-free equilibrium, producing prolonged intervals of near-zero prevalence interrupted by recurrent high-prevalence outbreaks (see green-to-orange trajectories in \autoref{fig:Figure_7}f).

In summary, the interaction of seasonality, behavioral feedback, and pathogen cooperation generates a wide spectrum of dynamical regimes in the SIRSsm coinfection model. Cooperation expands the parameter space supporting complex behaviors such as period-doubling cascades and chaos, and introduces additional mechanisms for transitions between endemic and disease-free states. These findings show the essential role of pathogen interactions in shaping epidemic patterns beyond single-pathogen dynamics.

\section{Discussion}
In this manuscript, we analyzed how seasonality and behavioral feedback affect the dynamics of multi-pathogen systems, especially with cooperative interactions, in which infection with one pathogen facilitates subsequent infection with the other \cite{PaperCoopSyndemics2013,cai2015avalanche}. Previous research has shown that cooperation can generate abrupt changes in the long-term prevalence of disease \cite{PaperCoopSyndemics2013,ExactSXP_Paper}. These sudden transitions indicate a risk of surprise outbreaks and imply that control rules based solely on single-disease models may miss key information needed to predict and eradicate disease \cite{kramer2025limitations,hebert2025one}. Our analysis proceeded in three stages, from the coinfection model without either mechanism, through each mechanism individually, to the full combined system.

The SIRS co-infection model exhibits a clearer version of the phase transition reported for the SIR-SIR case \cite{PaperCoopSyndemics2013,ExactSXP_Paper}. In our case, it arises from a backward bifurcation for $C>2$ in the region where both DFE and EE are stable. This phenomenon also appears in other models with self-boosting feedback strong enough to outweigh the self-inhibition from susceptible depletion \cite{aguiar2024backward,steindorf2022backward}.

Introducing additional time scales leads to qualitatively different long-term dynamics in the two limiting cases (pure seasonality vs. pure behavioral feedback), traced back to the mechanisms driving oscillations in $R_t$. Seasonality induces oscillations at a slow, fixed rate, independent of the phase-space variables, whereas mitigation drives oscillations in response to the history of the phase-space variable $\X$, if the maximum mitigation $m_{\rm max}$ is strong enough. As a result, mitigation-driven oscillations in the spreading rate can match the time scale of rising case numbers, producing low-incidence limit cycles or a stable high-prevalence fixed point. When seasonality and behavioral feedback have comparable strength, the system can exhibit limit cycles alternating between low and high prevalence every second year, as well as period-doubling cascades to chaos, as reported in single-pathogen models \cite{wagner2025societal, d2022behavioral}. Cooperativity among pathogens introduces another source of complexity, leading to coexisting attractors. For fast disease dynamics (low $\kappa$) and intermediate mitigation strengths, we observe a period-doubling cascade in the cooperativity parameter. This cascade interacts with the period-doubling behavior arising from the interplay of mitigation and seasonality at low $C$ values, previously reported in single-disease models. The lower maxima of the cooperativity-induced bifurcations appear to approach the DFE asymptotically, giving rise to years of near-zero prevalence within recurring endemic dynamics. For stronger cooperations, a third cascade emerges at high $m_{\rm max}$, which eventually merges with the cooperativity-induced cascade, forming a structure resembling an Arnold tongue. Comparable regimes have been documented in real multi-strain and multi-pathogen systems: in dengue, the interaction of serotypes under seasonal forcing produces low-amplitude limit cycles and chaotic attractors consistent with incidence records \cite{aguiar2011role}, and similar multi-annual patterns are known from seasonally forced childhood infections \cite{stollenwerk2022seasonally}.

Our approach, however, has some limitations. First, diseases in real settings are rarely symmetric in their initial conditions, mechanisms, and parameters \cite{kramer2025limitations}. For example, an emerging infectious disease may alter the prevalence of endemic sympatric diseases through interventions aimed at containing it (e.g., COVID-19 with the respiratory syncytial virus, RSV), or may differ in key characteristics (e.g., baseline spreading or recovery rates, or additional mechanisms). However, symmetry can be a reasonable assumption for drifting strains of a pathogen or for endemic diseases that share the same seasonal cycles. Recent studies have started to relax this assumption, showing that when two diseases interact asymmetrically (e.g., spreading on different time scales or with unequal coupling strengths), the faster or more strongly coupled pathogen can dominate the joint dynamics and shift the location of the tipping point \cite{ventura2021role,kramer2025limitations}. A documented example is the interaction between influenza and invasive group A \textit{Streptococcus} (iGAS): the incidence of iGAS increases more than 30-fold in the weeks following an influenza infection, with no comparable effect in the opposite direction \cite{goldsmith2024associations}. We speculate that breaking the symmetry in our framework would have several consequences: unequal cooperativities ($C_{AB}\neq C_{BA}$) could split the single first-order transition into two staggered ones, distinct disease time scales ($\kappa_A\neq\kappa_B$) could introduce a second, incommensurate frequency and favor quasiperiodic dynamics over period-doubling, and mitigation or seasonality acting unequally on the two diseases could desynchronize their waves into out-of-phase multi-annual patterns. We leave the systematic exploration of these asymmetric regimes for future work. Second, we treat the cooperativity $C$ as a constant multiplicative factor and probe large values ($C=15,~25$) to expose the full range of accessible dynamics, whereas physiological facilitation between pathogens is bounded and should saturate at high co-infection pressure; these large values are therefore a theoretical exploration rather than empirical estimates. The central mechanisms, however, do not depend on them: the backward bifurcation and the cooperativity-induced period-doublings already emerge at moderate $C\gtrsim2$ (see \autoref{fig:Figure_6} and \autoref{fig:Figure_7}). Furthermore, strong facilitation is not implausible. At the population level, the incidence of invasive group A streptococcal disease increases roughly 34-fold following an influenza infection \cite{goldsmith2024associations}, consistent with the synergy between influenza and respiratory bacteria \cite{mccullers2014copathogenesis,klein2016frequency}. Such incidence ratios combine enhanced acquisition with enhanced progression to disease, and thus overestimate the pure transmission boost, but they show that co-infection can raise the effective force of infection by more than an order of magnitude. In this sense, our model provides a general framework for anticipating the dynamics of newly emerging strains or pathogen pairs, which may fall in so-far unexplored regions of the parameter space. Third, we assume that the characteristic delay determining the behavioral feedback is static. Its value, however, was chosen so that we observe oscillations over a wide range of $m_{\rm max}$, as thoroughly studied in \cite{wagner2025societal}. Fourth, the models we use (ODE-based mean-field models) do not capture the stochasticity inherent to disease spread. Phenomena such as noise-induced tipping can lead to threshold crossings even below the deterministic critical transmission rate. Nonetheless, our goal is not to pinpoint the exact timing of transmission events, but rather to characterize the landscape of possible dynamics across the wider parameter space.

Natural extensions of our model should exploit heterogeneity that breaks the assumed symmetry. Straightforward modifications include allowing behavioral feedback to affect the two diseases differently, capturing differences in risk perception and transmission modes (e.g., droplets versus aerosols), or incorporating differential severity between the diseases. This would allow studying, for instance, how asymptomatic carriers of one disease might accelerate the dynamics of another due to strong cooperation.

Overall, our results show that pathogen cooperation must be explicitly accounted for, as it fundamentally alters epidemic dynamics. Understanding these interacting mechanisms is therefore essential for developing robust mitigation strategies that remain effective in the presence of cooperating pathogens and under changing spreading conditions.

\section*{Author Contributions}

Conceptualization: SC\\
Formal analysis: RAL\\
Investigation: RAL\\
Methodology: RAL, FG, SC\\
Software: RAL\\
Supervision: SC\\
Validation: RAL, FG, SC\\
Visualization: RAL, SC\\
Writing - Original Draft: RAL, SC\\
Writing - Review \& Editing: RAL, FG, SC\\

\section*{Code availability}

All code to reproduce the analysis and figures shown in the manuscript, as well as in the Supplementary Material, is available online on Github \href{https://github.com/rodrigolind/ComplexDynamics_cooperativeSIRSco-infectionModel}{https://github.com/rodrigolind/ComplexDynamics\_cooperativeSIRSco-infectionModel}.

\section*{Acknowledgments}

We thank the Priesemann group for fruitful discussions. Claude, Gemini PRO, Chat GPT-5.2, and Grammarly AI were used for grammar checks in the main text, and GitHub Copilot served as a coding assistant. The authors reviewed and assume full responsibility for the final content of the article.


\begin{thebibliography}{10}

\bibitem{d2022behavioral}
d'Onofrio A, Manfredi P.
\newblock Behavioral SIR models with incidence-based social-distancing.
\newblock Chaos, Solitons \& Fractals. 2022;159:112072.
\newblock Available from: \url{https://www.sciencedirect.com/science/article/pii/S096007792200282X}.

\bibitem{wagner2025societal}
Wagner J, Bauer S, Contreras S, Fleddermann L, Parlitz U, Priesemann V.
\newblock Societal self-regulation induces complex infection dynamics and chaos.
\newblock Physical Review Research. 2025;7(1):013308.

\bibitem{manfredi2013modeling}
Manfredi P, D'Onofrio A.
\newblock Modeling the interplay between human behavior and the spread of infectious diseases.
\newblock Springer Science \& Business Media; 2013.

\bibitem{d2009information}
d’Onofrio A, Manfredi P.
\newblock Information-related changes in contact patterns may trigger oscillations in the endemic prevalence of infectious diseases.
\newblock Journal of Theoretical Biology. 2009;256(3):473-8.

\bibitem{donges2022interplay}
D{\"o}nges P, Wagner J, Contreras S, Iftekhar EN, Bauer S, Mohr SB, et~al.
\newblock Interplay between risk perception, behavior, and COVID-19 spread.
\newblock Frontiers in Physics. 2022;10:842180.

\bibitem{fisman2007seasonality}
Fisman DN.
\newblock Seasonality of infectious diseases.
\newblock Annual review of public health. 2007;28(1):127-43.

\bibitem{pascual2005seasonal}
Pascual M, Dobson A.
\newblock Seasonal patterns of infectious diseases.
\newblock PLoS Medicine. 2005;2(1):e5.

\bibitem{altizer2006seasonality}
Altizer S, Dobson A, Hosseini P, Hudson P, Pascual M, Rohani P.
\newblock Seasonality and the dynamics of infectious diseases.
\newblock Ecology letters. 2006;9(4):467-84.

\bibitem{stollenwerk2022seasonally}
Stollenwerk N, Spaziani S, Mar J, Arrizabalaga IE, Knopoff D, Cusimano N, et~al.
\newblock Seasonally forced sir systems applied to respiratory infectious diseases, bifurcations, and chaos.
\newblock Computational and Mathematical Methods. 2022;2022(1):3556043.

\bibitem{contreras2023emergency}
Contreras S, Iftekhar EN, Priesemann V.
\newblock From emergency response to long-term management: the many faces of the endemic state of COVID-19.
\newblock The Lancet Regional Health--Europe. 2023;30.

\bibitem{siettos2013mathematical}
Siettos CI, Russo L.
\newblock Mathematical modeling of infectious disease dynamics.
\newblock Virulence. 2013;4(4):295-306.

\bibitem{shaw2025co}
Shaw KE, Peterson JK, Jalali N, Ratnavale S, Alkuzweny M, Barbera C, et~al.
\newblock Co-circulating pathogens of humans: a systematic review of mechanistic transmission models.
\newblock Proceedings of the Royal Society B: Biological Sciences. 2025;292(2055).

\bibitem{rohani2003ecological}
Rohani P, Green C, Mantilla-Beniers N, Grenfell BT.
\newblock Ecological interference between fatal diseases.
\newblock Nature. 2003;422(6934):885-8.

\bibitem{contreras2023model}
Contreras S, Or{\'o}stica KY, Daza-Sanchez A, Wagner J, D{\"o}nges P, Medina-Ortiz D, et~al.
\newblock Model-based assessment of sampling protocols for infectious disease genomic surveillance.
\newblock Chaos, Solitons \& Fractals. 2023;167:113093.

\bibitem{orostica2024early}
Or{\'o}stica KY, Mohr SB, Dehning J, Bauer S, Medina-Ortiz D, Iftekhar EN, et~al.
\newblock Early mutational signatures and transmissibility of SARS-CoV-2 Gamma and Lambda variants in Chile.
\newblock Scientific Reports. 2024;14(1):16000.

\bibitem{Cooperationbtw_Infl_StrepPneu}
McCullers JA, Rehg JE.
\newblock Lethal Synergism between Influenza Virus and Streptococcus pneumoniae: Characterization of a Mouse Model and the Role of Platelet-Activating Factor Receptor.
\newblock The Journal of Infectious Diseases. 2002 08;186(3):341-50.
\newblock Available from: \url{https://doi.org/10.1086/341462}.

\bibitem{mccullers2014copathogenesis}
McCullers JA.
\newblock The co-pathogenesis of influenza viruses with bacteria in the lung.
\newblock Nature Reviews Microbiology. 2014;12(4):252-62.

\bibitem{morris2017secondary}
Morris DE, Cleary DW, Clarke SC.
\newblock Secondary bacterial infections associated with influenza pandemics.
\newblock Frontiers in microbiology. 2017;8:1041.

\bibitem{Cooperationbtw_HIV_HCV_A}
Deng A, Chen C, Ishizaka Y, Chen X, Sun B, Yang R.
\newblock Human immunodeficiency virus type 1 Vpr increases hepatitis C virus RNA replication in cell culture.
\newblock Virus Research. 2014;184:93-102.
\newblock Available from: \url{https://www.sciencedirect.com/science/article/pii/S0168170214000811}.

\bibitem{Cooperationbtw_HIV_HCV_B}
Chen JY, Feeney ER, Chung RT.
\newblock HCV and HIV co-infection: mechanisms and management.
\newblock Nature Reviews Gastroenterology \& Hepatology. 2014 Jun;11(6):362-71.
\newblock Available from: \url{https://doi.org/10.1038/nrgastro.2014.17}.

\bibitem{Cooperationbtw_HIV_HCV_C}
Gobran ST, Ancuta P, Shoukry NH.
\newblock A Tale of Two Viruses: Immunological Insights Into HCV/HIV Coinfection.
\newblock Frontiers in Immunology. 2021;Volume 12 - 2021.
\newblock Available from: \url{https://www.frontiersin.org/journals/immunology/articles/10.3389/fimmu.2021.726419}.

\bibitem{Cooperationbtw_PapillomaVirusTypes}
Liao G, Jiang X, She B, Tang H, Wang Z, Zhou H, et~al.
\newblock Multi-Infection Patterns and Co-infection Preference of 27 Human Papillomavirus Types Among 137,943 Gynecological Outpatients Across China.
\newblock Frontiers in Oncology. 2020;10:449.
\newblock Available from: \url{https://doi.org/10.3389/fonc.2020.00449}.

\bibitem{CoInfectionNumbers_SARS-CoV-2}
Hu X, Zhang F, Jia J, Xin X, Dai X, Dong L, et~al.
\newblock Associated factors of respiratory co-infection of COVID-19 and the impact of co-infection on SARS-CoV-2 viral load.
\newblock Journal of Infection in Developing Countries. 2024 Aug;18(8):1204-11.
\newblock Epub ahead of print.

\bibitem{CoInfectionNumbers_SARS-CoV-2_mortality}
Musuuza JS, Watson L, Parmasad V, Putman-Buehler N, Christensen L, Safdar N.
\newblock Prevalence and outcomes of co-infection and superinfection with SARS-CoV-2 and other pathogens: A systematic review and meta-analysis.
\newblock PLoS ONE. 2021;16(5):e0251170.
\newblock Available from: \url{https://doi.org/10.1371/journal.pone.0251170}.

\bibitem{Cooperationbtw_Infl_IPD}
Berry I, Tuite AR, Salomon A, Drews S, Harris AD, Hatchette T, et~al.
\newblock Association of Influenza Activity and Environmental Conditions With the Risk of Invasive Pneumococcal Disease.
\newblock JAMA Network Open. 2020 07;3(7):e2010167-7.
\newblock Available from: \url{https://doi.org/10.1001/jamanetworkopen.2020.10167}.

\bibitem{klugman2009deadly}
Klugman KP, Chien YW, Madhi SA.
\newblock Pneumococcal pneumonia and influenza: a deadly combination.
\newblock Vaccine. 2009;27:C9-C14.

\bibitem{messiah2012risk}
Messiah A, Constant A, Contrand B, Felonneau ML, Lagarde E.
\newblock Risk compensation: a male phenomenon? Results from a controlled intervention trial promoting helmet use among cyclists.
\newblock American journal of public health. 2012;102(S2):S204-6.

\bibitem{mueller2025paradox}
Müller L, Mallick P, Marín-Carballo AB, Dönges P, Kettlitz RJN, Klett-Tammen CJ, et~al.
\newblock Testing paradox may explain increased observed prevalence of bacterial {STIs} among {MSM} on {HIV PrEP}: A modeling study.
\newblock Proceedings of the National Academy of Sciences. 2025;122(44):e2524944122.
\newblock Available from: \url{https://www.pnas.org/doi/abs/10.1073/pnas.2524944122}.

\bibitem{mallick2026stability}
Mallick P, M{\"u}ller L, Mar{\'\i}n-Carballo AB, D{\"o}nges P, Contreras S.
\newblock Stability and bifurcations of a minimal model for the effect of PrEP-related risk compensation in epidemics of sexually transmitted infections.
\newblock Chaos, Solitons \& Fractals. 2026;210:118649.

\bibitem{steindorf2022backward}
Steindorf V, Srivastav AK, Stollenwerk N, Kooi BW, Aguiar M.
\newblock Modeling secondary infections with temporary immunity and disease enhancement factor: Mechanisms for complex dynamics in simple epidemiological models.
\newblock Chaos, Solitons \& Fractals. 2022;164:112709.
\newblock Available from: \url{https://www.sciencedirect.com/science/article/pii/S0960077922008888}.

\bibitem{aguiar2024backward}
Aguiar M, Steindorf V, Srivastav AK, Stollenwerk N, Kooi BW.
\newblock Bifurcation analysis of a two infection SIR-SIR epidemic model with temporary immunity and disease enhancement.
\newblock Nonlinear Dynamics. 2024;112(15):13621-39.
\newblock Available from: \url{https://doi.org/10.1007/s11071-024-09710-9}.

\bibitem{kramer2025limitations}
Kramer SC, Pirikahu S, Kussmaul C, Opatowski L, Domenech~de Cell{\`e}s M.
\newblock Limitations of non-mechanistic methods for characterizing pathogen-pathogen interactions: A simulation study.
\newblock bioRxiv. 2025:2025-12.

\bibitem{PaperCoopSyndemics2013}
Chen L, Ghanbarnejad F, Cai W, Grassberger P.
\newblock Outbreaks of coinfections: The critical role of cooperativity.
\newblock EPL (Europhysics Letters). 2013 12;104.

\bibitem{CoopSyndemics_Khazaee_2022}
Khazaee A, Ghanbarnejad F.
\newblock Effects of measures on phase transitions in two cooperative susceptible-infectious-recovered dynamics.
\newblock Physical Review E. 2022 Mar;105(3).
\newblock Available from: \url{http://dx.doi.org/10.1103/PhysRevE.105.034311}.

\bibitem{ExactSXP_Paper}
Zarei F, Moghimi-Araghi S, Ghanbarnejad F.
\newblock Exact solution of generalized cooperative susceptible-infected-removed (SIR) dynamics.
\newblock Phys Rev E. 2019 Jul;100:012307.
\newblock Available from: \url{https://link.aps.org/doi/10.1103/PhysRevE.100.012307}.

\bibitem{BckwBif_DueToExReinf3}
Feng Z, Castillo-Chavez C, Capurro AF.
\newblock A Model for Tuberculosis with Exogenous Reinfection.
\newblock Theoretical Population Biology. 2000;57(3):235-47.
\newblock Available from: \url{https://www.sciencedirect.com/science/article/pii/S0040580900914515}.

\bibitem{BckwBif_DueToExReinf1}
Cohen T, Colijn C, Finklea B, Murray M.
\newblock Exogenous re-infection and the dynamics of tuberculosis epidemics: local effects in a network model of transmission.
\newblock Journal of The Royal Society Interface. 2007;4(14):523-31.
\newblock Available from: \url{https://royalsocietypublishing.org/doi/abs/10.1098/rsif.2006.0193}.

\bibitem{BckwBif_DueToExReinf2}
Bhunu CP.
\newblock Mathematical analysis of a three-strain tuberculosis transmission model.
\newblock Applied Mathematical Modelling. 2011;35(9):4647-60.
\newblock Available from: \url{https://www.sciencedirect.com/science/article/pii/S0307904X11001739}.

\bibitem{BckwBif_DueToSuperInfPapilloma}
Reluga TC, Medlock J, Perelson AS.
\newblock Backward bifurcations and multiple equilibria in epidemic models with structured immunity.
\newblock Journal of Theoretical Biology. 2008;252(1):155-65.
\newblock Available from: \url{https://www.sciencedirect.com/science/article/pii/S0022519308000271}.

\bibitem{BckwBif_DueToRelapse}
Wangari IM, Stone L.
\newblock Backward bifurcation and hysteresis in models of recurrent tuberculosis.
\newblock PLoS ONE. 2018;13(3):e0194256.
\newblock ECollection 2018.

\bibitem{BckwBif_DueToVacc1}
Brauer F.
\newblock Backward bifurcations in simple vaccination models.
\newblock Journal of Mathematical Analysis and Applications. 2004;298(2):418-31.
\newblock Available from: \url{https://www.sciencedirect.com/science/article/pii/S0022247X04004378}.

\bibitem{BckwBif_DueToVacc2}
Elbasha EH, Gumel AB.
\newblock Theoretical assessment of public health impact of imperfect prophylactic HIV-1 vaccines with therapeutic benefits.
\newblock Bulletin of Mathematical Biology. 2006 Apr;68(3):577-614.
\newblock Epub 2006 Apr 7.

\bibitem{BckwBif_DueToVacc3}
Kribs-Zaleta CM, Velasco-Hernández JX.
\newblock A simple vaccination model with multiple endemic states.
\newblock Mathematical Biosciences. 2000;164(2):183-201.
\newblock Available from: \url{https://www.sciencedirect.com/science/article/pii/S0025556400000031}.

\bibitem{BackwardsBif_CausesandExamples}
Gumel AB.
\newblock Causes of backward bifurcations in some epidemiological models.
\newblock Journal of Mathematical Analysis and Applications. 2012;395(1):355-65.
\newblock Available from: \url{https://www.sciencedirect.com/science/article/pii/S0022247X12003551}.

\bibitem{NextGenMatrix}
{van den Driessche} P, Watmough J.
\newblock Reproduction numbers and sub-threshold endemic equilibria for compartmental models of disease transmission.
\newblock Mathematical Biosciences. 2002;180(1):29-48.
\newblock Available from: \url{https://www.sciencedirect.com/science/article/pii/S0025556402001086}.

\bibitem{OnR0Def}
Diekmann O, Heesterbeek JAP, Metz JAJ.
\newblock On the definition and the computation of the basic reproduction ratio R0 in models for infectious diseases in heterogeneous populations.
\newblock Journal of Mathematical Biology. 1990;28(4):365-82.

\bibitem{InclBeh_Onofrio_Old_Vacc_ErlangKernel}
d’Onofrio A, Manfredi P, Salinelli E.
\newblock Vaccinating behaviour, information, and the dynamics of SIR vaccine preventable diseases.
\newblock Theoretical Population Biology. 2007;71(3):301-17.
\newblock Available from: \url{https://www.sciencedirect.com/science/article/pii/S0040580907000020}.

\bibitem{DescartesRuleOfSigns_OldPaper}
Curtiss DR.
\newblock Recent Extentions of Descartes' Rule of Signs.
\newblock Annals of Mathematics. 1918;19(4):251-78.
\newblock Available from: \url{http://www.jstor.org/stable/1967494}.

\bibitem{DescartesRuleOfSigns_Proof}
Wang X.
\newblock A Simple Proof of Descartes's Rule of Signs.
\newblock American Mathematical Monthly. 2004 06;111.

\bibitem{Chen2017Fundamental}
Chen L, Ghanbarnejad F, Brockmann D.
\newblock Fundamental properties of cooperative contagion processes.
\newblock New Journal of Physics. 2017 nov;19(10):103041.
\newblock Available from: \url{https://doi.org/10.1088/1367-2630/aa8bd2}.

\bibitem{cai2015avalanche}
Cai W, Chen L, Ghanbarnejad F, Grassberger P.
\newblock Avalanche outbreaks emerging in cooperative contagions.
\newblock Nature physics. 2015;11(11):936-40.

\bibitem{hebert2025one}
H{\'e}bert-Dufresne L, Ahn YY, Allard A, Colizza V, Crothers JW, Dodds PS, et~al.
\newblock One pathogen does not an epidemic make: a review of interacting contagions, diseases, beliefs, and stories.
\newblock npj Complexity. 2025;2(1):26.

\bibitem{aguiar2011role}
Aguiar M, Ballesteros S, Kooi BW, Stollenwerk N.
\newblock The role of seasonality and import in a minimalistic multi-strain dengue model capturing differences between primary and secondary infections: complex dynamics and its implications for data analysis.
\newblock Journal of theoretical biology. 2011;289:181-96.

\bibitem{ventura2021role}
Ventura PC, Moreno Y, Rodrigues FA.
\newblock Role of time scale in the spreading of asymmetrically interacting diseases.
\newblock Physical Review Research. 2021;3(1):013146.

\bibitem{goldsmith2024associations}
Goldsmith JJ, Vu C, Zhu Z, MacLachlan JH, Thomson TN, Campbell PT, et~al.
\newblock The associations between invasive group A streptococcal disease and infection with influenza, varicella, or hepatitis C viruses: a data linkage study, Victoria, Australia.
\newblock International Journal of Infectious Diseases. 2024;141:106969.

\bibitem{klein2016frequency}
Klein EY, Monteforte B, Gupta A, Jiang W, May L, Hsieh YH, et~al.
\newblock The frequency of influenza and bacterial coinfection: a systematic review and meta-analysis.
\newblock Influenza and other respiratory viruses. 2016;10(5):394-403.

\end{thebibliography}

\clearpage

\newpage
\renewcommand{\thefigure}{S\arabic{figure}}
\renewcommand{\figurename}{Supplementary~Figure}
\setcounter{figure}{0}
\renewcommand{\thetable}{S\arabic{table}}
\renewcommand{\tablename}{Supplementary~Table}

\setcounter{table}{0}
\renewcommand{\thesection}{S\arabic{section}}
\setcounter{section}{0}
\setcounter{page}{1}
\section{Supplementary Material}

\subsection{Basic reproduction number}
\label{sec: appendix R0 calc}

The basic reproduction number is defined as the expected number of secondary cases produced, in a completely susceptible population, by a typical infective individual \cite{OnR0Def}. We calculate the basic reproduction number by the method of \textit{the next generation matrix} \cite{NextGenMatrix,OnR0Def}. To this end we first rewrite the ODE equations of the infected compartments \eqref{eq:SIRSsm_first infected}-\eqref{eq:SIRSsm_last infected} as the ODE of the single vector $\mathcal X_{I}:=(\A,\B,\AB,\Ab,\aB)^T$ containing all of the phase space variables corresponding to infected disease states. We define $\mathcal F$ as the summands in the ODE system that contribute to new infection cases and $\mathcal V$ as the vector containing all the other fluxes multiplied by $-1$. The ODE system for the infected compartments can then be written as
\begin{align}
    \dot{\mathcal X_{I}}=\mathcal F-\mathcal V,
\end{align}
where
\begin{align}
    \mathcal F=\begin{pmatrix}
        \alpha\XA S \\ \alpha\XB S \\ C \alpha (\XB \A+\XA \B) \\ C \alpha \XA \bb \\ C \alpha\XB\aVar
    \end{pmatrix} \text{ and }
    \mathcal V=\begin{pmatrix}
        C \alpha\XB\A+\gamma\A-\nu\Ab \\ C \alpha\XA\B+\gamma\B-\nu\aB \\ 2\gamma\AB \\ \gamma(\Ab-\AB)+\nu\Ab \\ \gamma(\aB-\AB)+\nu\aB
    \end{pmatrix}.
\end{align}

The jacobians with respect to each of the vectors $\mathcal F$ and $\mathcal V$ at a disease-free state yield
\begin{align}
    \mathcal {D_{X_{I}}F}=\begin{pmatrix}
        \alpha S &0&\alpha S&\alpha S&0 \\
        0&\alpha S&\alpha S&0&\alpha S \\
        C \alpha (\XB + \B)&C \alpha (\A + \XA)&C \alpha (\A + \B)&C \alpha\B&C \alpha\A \\
        C \alpha\bb&0&C \alpha\bb&C \alpha\bb&0 \\
        0&C \alpha\aVar&C \alpha\aVar&0&C \alpha\aVar
    \end{pmatrix}\stackrel{DFE}{=}\begin{pmatrix}
        \alpha &0&\alpha &\alpha &0 \\
        0&\alpha &\alpha &0&\alpha  \\
        0&0&0&0&0 \\
        0&0&0&0&0 \\
        0&0&0&0&0
    \end{pmatrix}\\ \text{ and }
    \mathcal {D_{X_{I}}V}=\begin{pmatrix}
        C \alpha\XB+\gamma & C \alpha\A & C \alpha\A&-\nu&C \alpha\A \\
        C \alpha\B&C \alpha\XA+\gamma&C \alpha\B&C \alpha\B&-\nu\\
        0&0&2\gamma&0&0 \\
        0&0&-\gamma&\gamma+\nu&0 \\
        0&0&-\gamma&0&\gamma+\nu
    \end{pmatrix}\stackrel{DFE}{=}\begin{pmatrix}
        \gamma&0&0&-\nu &0 \\
        0&\gamma&0&0&-\nu  \\
        0&0&2\gamma&0&0 \\
        0&0&-\gamma&\gamma+\nu&0 \\
        0&0&-\gamma&0&\gamma+\nu
    \end{pmatrix}.
\end{align}

With the inverse
\begin{align}
    (\mathcal {D_{X_{I}}V}\big|_{\rm DFE})^{-1}=\begin{pmatrix}
        \frac{1}{\gamma}&0&\frac{\nu}{2\gamma(\gamma+\nu)}&\frac{2\nu}{2\gamma(\gamma+\nu)}&0\\
        0&\frac{1}{\gamma}&\frac{\nu}{2\gamma(\gamma+\nu)}&0&\frac{2\nu}{2\gamma(\gamma+\nu)}\\
        0&0&\frac{\gamma+\nu}{2\gamma(\gamma+\nu)}&0&0\\
        0&0&\frac{\gamma}{2\gamma(\gamma+\nu)}&\frac{2\gamma}{2\gamma(\gamma+\nu)}&0\\
        0&0&\frac{\gamma}{2\gamma(\gamma+\nu)}&0&\frac{2\gamma}{2\gamma(\gamma+\nu)}
    \end{pmatrix},
\end{align}
we can calculate the ratio of the jacobians $\mathcal K_{ij}$
\begin{align}
    \mathcal K=(\mathcal {D_{X_{I}}F}\big|_{\rm DFE})
    (\mathcal {D_{X_{I}}V}\big|_{\rm DFE})^{-1}=\frac{\alpha}{\gamma}\begin{pmatrix}
        1&0&1&1&0\\
        0&1&1&0&1\\
        0&0&0&0&0\\
        0&0&0&0&0\\
        0&0&0&0&0\\
    \end{pmatrix},
\end{align}
representing the expected number of new infections in compartment $i$ (rows) caused by an individual initially in compartment $j$ (columns) in a fully susceptible population. The basic reproduction number is then defined as the spectral radius (largest eigenvalue) of $\mathcal K$
\begin{align}
    R_0\equiv\rho(\mathcal K)=\frac{\alpha}{\gamma}.
\end{align}
Note that the basic reproduction number is independent of $\beta$ and, in fact, identical to that of the corresponding single-disease models. This is because $\beta$ acts on the infected compartment $P$, which vanishes in a fully susceptible population and therefore does not contribute to $R_0$.

\subsection{Derivation of the equilibria}
\label{sec: appendix expl FP calc}

As discussed in \autoref{sec: the system}, the proposed SIRS co-infection model can be written in reduced form with the variables $S,P,\X,Q,\A$ and $\AB$. The fixed points are obtained by setting all the time derivatives of the phase-space variables to 0

\begin{align}
\label{App: eq:SIRS_first}
    0=\dot S^*&=-2\alpha \X^*\cdot S^*+2\nu (P^*-\A^*)\\
\label{App: eq:SIRS_dot X}
    0=\dot \X^*&=+\alpha \X^*\cdot S^*+C \alpha \X^*\cdot P^*-\gamma \X^*\\
\label{App: eq:SIRS_dot P}
    0=\dot P^* &=+\alpha \X^*\cdot S^*-C \alpha \X^*\cdot P^*+\nu(\X^*-P^*+Q^*-2\AB^*)\\
\label{App: eq:SIRS_dot Q}
    0=\dot Q^*&=2C \alpha \X^* \A^*+2\gamma(-2\AB^*+\X^*-\A^*)-2\nu(Q^*-\AB^*)\\
\label{App: eq:SIRS_dot A}
    0=\dot \A^*&=\alpha \X^* S^*-C \alpha \X^* \A^*-\gamma \A^*+\nu(\X^*-\AB^*-\A^*)\\
\label{App: eq:SIRS_last}
    0=\dot{\AB^*}&=2C \alpha \X^* \A^*-2\gamma \AB^*.
\end{align}

As the six equations we have are linearly dependent we can only eliminate 5 of the 6 variables with them. The last variable of each solution is fixed by the normalization constraint \eqref{eq: total pop=1, SIRS, SXPQAAB-Var}.

Firstly, we can identify the disease-free equilibrium (DFE) $\X^*=0$ as a solution to \eqref{App: eq:SIRS_dot X}. From the other equations, we can quickly see that, in this case, all of the other variables, excluding $S^*$, also vanish. Using the normalization constraint \eqref{eq: total pop=1, SIRS, SXPQAAB-Var}, we obtain:
\begin{align}
\label{App: eq: DFE}
(S^*,\X^*,P^*,Q^*,\A^*,\AB^*) = (1,0,0,0,0,0).
\end{align}

Considering other solutions besides the DFE, we can divide by $\X^*\neq0$, yielding
\begin{align}
    \label{App: eq:SIRS_dot X /X}
    0\stackrel{\eqref{App: eq:SIRS_dot X}}{=}+\alpha S^*+C \alpha P^*-\gamma.
\end{align}

As a first step, we write the variables $S^*,Q^*,A^*$ and $AB^*$ in terms of $\X^*$ and $P^*$ (using four of the six equations). We get

\begin{align}
\label{App: eq: S(P,X)}
    S^*&\stackrel{\eqref{App: eq:SIRS_dot X /X}}{=}\frac{1}{\alpha}(\gamma-C \alpha P^*)\\
\label{App: eq: A(P,X)}
    \A^*&\stackrel{\eqref{App: eq:SIRS_first}\eqref{App: eq: S(P,X)}}{=}P^*+\frac{1}{\nu}(C \alpha P^*-\gamma)\X^*\\
\label{App: eq: AB(P,X)}
    \AB^*&\stackrel{\eqref{App: eq:SIRS_last}\eqref{App: eq: A(P,X)}}{=}\frac{C \alpha}{\gamma}\X^* P^*+\frac{C \alpha}{\gamma\nu}(C \alpha P^*-\gamma)\X^{*^2}\\
\label{App: eq: Q(P,X)}
\begin{split}
    Q^*&\stackrel{\eqref{App: eq:SIRS_dot P}\eqref{App: eq: S(P,X)}}{=}\frac{1}{\nu}(2C \alpha P^*-\gamma)\X^*-\X^*+P^*+2\AB^*\\
    &\stackrel{\eqref{App: eq: AB(P,X)}}{=}P^*-\rho \X^*+2\frac{C \alpha\rho}{\gamma}\X^* P^*+\frac{C \alpha}{\gamma\nu}(C \alpha P^*-\gamma)\X^{*^2},
\end{split}
\end{align}

where we introduce the dimensionless parameter $\rho:=1+\frac{\gamma}{\nu}$ to alleviate the notation. Now we proceed to compute $P^*$ as a function of $\X^*$. From \eqref{App: eq:SIRS_dot A}, we get

\begin{equation}
    \X^*(\alpha S^*-C \alpha \A^*+\nu)\stackrel{\eqref{App: eq:SIRS_dot A}}{=}\gamma \A^*+\nu(\AB^*+\A^*),
\end{equation}
which after replacing it in \eqref{App: eq: S(P,X)}, \eqref{App: eq: A(P,X)} and \eqref{App: eq: AB(P,X)} become

\begin{multline}
\label{temp1}
    \X^*\left[\gamma-2C \alpha P^*-\frac{C \alpha}{\nu}\X^*(C \alpha P^*-\gamma)+\nu\right]=\frac{\gamma}{\nu}(C \alpha P^*-\gamma)\X^*+\frac{\nu C \alpha}{\gamma}\X^* P^*+\frac{C \alpha}{\gamma}(C \alpha P^*-\gamma)\X^{*^2}\\ +(\gamma+\nu) P^*+\X^*(C \alpha P^*-\gamma),
\end{multline}

\begin{align}
\label{temp2}
    \Leftrightarrow P^*\left[\frac{(C \alpha)^2\rho}{\gamma}\X^{*^2}+C \alpha \X^*+\frac{C \alpha\nu\rho^2}{\gamma}\X^*+\nu\rho\right]\stackrel{\eqref{temp1}}{=}C \alpha\rho \X^{*^2}+\nu\rho^2\X^*,
\end{align}
yielding
\begin{align}
\label{App: eq: P(X)}
    P^*\stackrel{\eqref{temp2}}{=}\frac{C \alpha\gamma\rho \X^{*^2}+\nu\gamma\rho^2 \X^*}{(C \alpha)^2\rho \X^{*^2}+C \alpha(\gamma+\nu\rho^2)\X^*+\nu\gamma\rho}.
\end{align}
Here, we assume the denominator to be non-zero. Finally, the only missing variable to compute is $\X^*$. Using the normalization constraint, we get

\begin{align}
\label{temp3}
\begin{split}
    1&\stackrel{\eqref{eq: total pop=1, SIRS, SXPQAAB-Var}}{=}S^*+2\X^*-2\AB^*+2P^*-2\A^*+Q^*\\
    &\stackrel{\eqref{App: eq: Q(P,X)}}{=}S^*+\left(1-\frac{\gamma}{\nu}\right)\X^*+\left(3+2\frac{C \alpha}{\nu}\X^*\right)P^*-2\A^*\\
    &\stackrel{\eqref{App: eq: S(P,X)}\eqref{App: eq: A(P,X)}}{=}\frac{\gamma}{\alpha}+\rho \X^*+\left(1-\frac{C \alpha}{\alpha}\right)P^*\\
    &\stackrel{\eqref{App: eq: P(X)}}{=}\frac{(C \alpha)^2\rho^2\X^{*^3}+\rho C \alpha(2\gamma+\nu\rho^2)\X^{*^2}+(2\nu\rho^2+C\gamma)\gamma \X^*+\frac{\nu\gamma^2\rho}{\alpha}}{(C \alpha)^2\rho \X^*+C \alpha(\gamma+\nu\rho^2)\X^*+\nu\gamma\rho}.
\end{split}
\end{align}

Thus, the fixed points for $\X^*$ obey the cubic equation
\begin{align}
\label{App: eq: CbEq for X}
    0\stackrel{\eqref{temp3}}{=}\underbrace{(C\alpha)^2\rho^2}_{:=a}\X^{*^3}+\underbrace{\rho C\alpha(2\gamma+\nu\rho^2-C\alpha)}_{:=b}\X^{*^2}+\underbrace{\left[C\alpha\left(\frac{\gamma^2}{\alpha}-\gamma-\nu\rho^2\right)+2\nu\gamma\rho^2\right]}_{:=c}\X^*+\underbrace{\left(\frac{\gamma}{\alpha}-1\right)\nu\gamma\rho}_{:=d}.
\end{align}
Hence, there are three solutions for $\X^*$, two of which will turn out to be the branches of the endemic equilibrium. The third will turn out to be negative and therefore unphysical.

Finally, in the case of the denominator in \eqref{App: eq: P(X)} vanishing we get
\begin{align}
\label{temp4}
    0=(C\alpha)^2\rho \X^{*^2}+C\alpha(\gamma+\nu\rho^2)\X^*+\nu\gamma\rho~~&\Leftrightarrow~~\X^*\in\left[-\frac{\nu\rho}{C\alpha},~-\frac{\gamma}{C\alpha\rho}\right],\\
    0\stackrel{\eqref{temp2}\eqref{temp4}}{=}C\alpha\rho \X^{*^2}+\nu\rho^2\X^*~~&\Leftrightarrow~~\X^*\in\left[-\frac{\nu\rho}{C\alpha},~0\right],
\end{align}
instead. The only solution solving both equations is $\X^*=-\frac{\nu\rho}{C\alpha}$, which can be neglected as it is negative and therefore unphysical.

\newpage

\begin{figure}
    \centering
    \includegraphics[width=88mm]{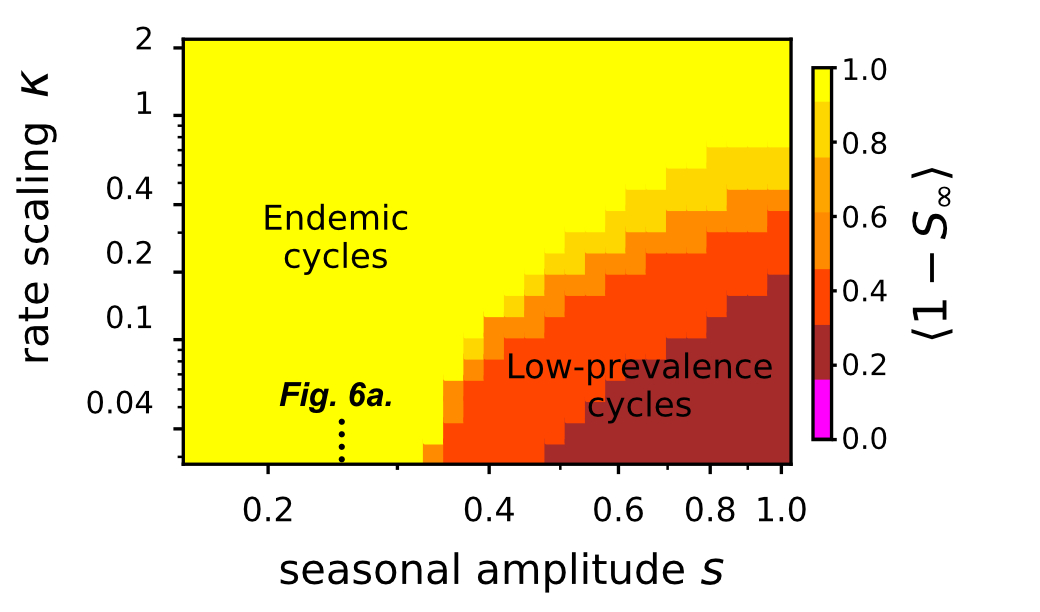}
    \caption{\textbf{Long-term dynamics of the SIRSsm coinfection model.} For $C=15$, dynamics are shown as a function of $\kappa$ and seasonal forcing $s$ (\textbf{a}, $m_{\rm max} = 0.2$). Across the parameter space, only limit cycles or more complex long-term behaviors are observed; fixed points are absent.
    Unspecified parameters use default values from \autoref{tab: Overview Parameters}.
    }
    \label{fig:Figure_S1}
\end{figure}

\end{document}